\pdfoutput=1
\documentclass[fleqn,usenatbib]{mnras}

\usepackage{newtxtext,newtxmath}

\usepackage[T1]{fontenc}
\usepackage{ae,aecompl}

\usepackage{graphicx}	
\usepackage{amsmath}	
\usepackage{amssymb}	
\hypersetup{draft}

\title[Sculptor Star Formation History]{The star formation history of the Sculptor dwarf spheroidal galaxy}

\author[M. Bettinelli et al.]{
M. Bettinelli,$^{1,2,3}$\thanks{E-mail: mbettine@iac.es}
S.~L. Hidalgo$^{1,2}$,
S.~Cassisi$^{4,1}$,
A.~Aparicio$^{2,1}$,
G.~Piotto$^{3,5}$,\newauthor
F.~Valdes$^{6}$,
A.~R.~Walker$^{7}$
\\
$^{1}$Instituto de Astrof\`isica de Canarias, V\`{i}a L\`{a}ctea S/N, E-38200 La Laguna, Tenerife, Spain\\
$^{2}$Department of Astrophysics, University of La Laguna, E-38200 La Laguna, Tenerife, Canary Islands, Spain\\
$^{3}$Dipartimento di Fisica e Astronomia ``Galileo Galilei'', Universit\`{a} degli Studi di Padova,  Vicolo dell'Osservatorio 3, I-35122 Padova, Italy\\
$^{4}$INAF-Osservatorio Astronomico d'Abruzzo, Via M. Maggini, I-64100 Teramo, Italy\\
$^{5}$INAF-Osservatorio Astronomico di Padova, Vicolo dell'Osservatorio 5, I-35122 Padova, Italy\\
$^{6}$National Optical Astronomy Observatory, P.O. Box 26732, Tucson, AZ 85719, USA\\
$^{7}$Cerro Tololo Inter-American Observatory, National Optical Astronomy Observatory, Casilla 603, La Serena, Chile\\
}

\date{Accepted XXX. Received YYY; in original form ZZZ}

\pubyear{2019}

\begin{document}
\label{firstpage}
\pagerange{\pageref{firstpage}--\pageref{lastpage}}
\maketitle

\begin{abstract}
We present the star formation history (SFH) of the Sculptor dwarf spheroidal galaxy based on deep $g$,$r$ photometry taken with DECam at the Blanco telescope, focusing our analysis on the central region of the galaxy extended up to $\sim3$ core radii.
We have investigated how the SFH changes radially, subdividing the sampled area into four regions, and have detected a clear trend of star formation. All the SFHs show a single episode of star formation, with the innermost region presenting a longer period of star formation of $\sim 1.5$ Gyr and for the outermost region the main period of star formation is confined to $\sim 0.5$ Gyr.
We observe a gradient in the mean age which is found to increase going towards the outer regions. 
These results suggest that Sculptor continued forming stars after the reionization epoch in its central part, while in the peripheral region the majority of stars probably formed during the reionization epoch and soon after its end. From our analysis Sculptor can not be considered strictly as a fossil of the reionization epoch.
\end{abstract}



\section{Introduction}
The first dwarf spheroidal (dSph) galaxy to be discovered early in the last century by \citet{1938BHarO.908....1S} is Sculptor. Since then, many other dwarf galaxies satellites of the MW have been discovered.
Historically, Sculptor, with Fornax, Leo I, Leo II, Draco, Ursa Minor, Carina, Sextans, is considered as the prototypical dwarf spheroidal galaxy \citep{2012AJ....144....4M}; these galaxies are defined {\it early-type dwarf galaxies} (\citealt{1999A&ARv...9..273V};\citealt{2009ARA&A..47..371T}) thus systems with a total luminosity of $M_{B}>-14$, characterized by low optical surface brightness ($V\sim22-26 \,\,{\rm mag/arcsec^{2}}$), no nucleus and poor gas content \citep{1994PASP..106.1225G}.
Despite these similarities, almost all dwarfs disclose their own peculiarities in the properties of their stellar populations such as in the SFH, chemical patterns, stellar variable populations and dark matter content.

Sculptor has been studied in depth since its discovery, photometrically, spectroscopically and in the radio wavelengths, in order to detect the presence of neutral hydrogen (HI).
Sculptor was found to be embedded in two HI clouds symmetrically distributed respect to the optical center (\citealt{1998AJ....116.1690C};\citealt{2003AJ....126.1295B};\citealt{2006AJ....131.1445P}). This finding is very important since Sculptor is the only dwarf spheroidal galaxy known to have retained this gas. The major parameters relative to this system are listed in Table~\ref{tab:param}.

Even though, at a first sight, the morphology of its CMD appears quite simple, the analysis of its stellar populations and SFH is not trivial, as will be shown in the present analysis. 
The published CMDs do not show a strong intermediate-age population, while the large RGB spread suggests the presence of internal age or abundances variations \citep{1984ApJ...285..483D}. 

\citet{1999PASP..111.1392M} observed a small region of $\sim 2$ arcmin$^{2}$ located $14.1$ arcmin from the center of Sculptor, using the Wide-Field Planetary Camera 2 (WFPC2) aboard the HST. The resulting CMD reaches $3$ magnitudes below the oldest MS turn-off and it resembles the stellar population of the earliest globular clusters.
In the same year, \citet{1999ApJ...523L..25H} published a wide field deep CMD based on data taken with Big Throughput Camera (BTC) at the CTIO $4$ m telescope. The quality of this CMD, in terms of its combination of depth and area, is unprecedented.

\begin{figure}

\includegraphics[width=\columnwidth]{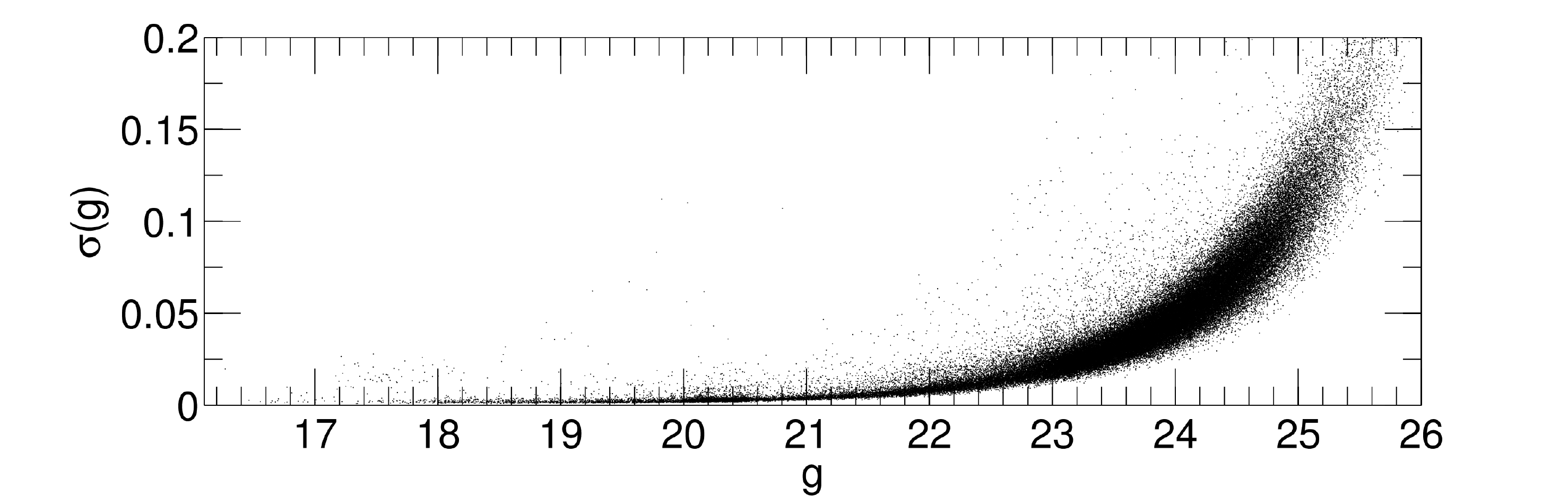}
\includegraphics[width=\columnwidth]{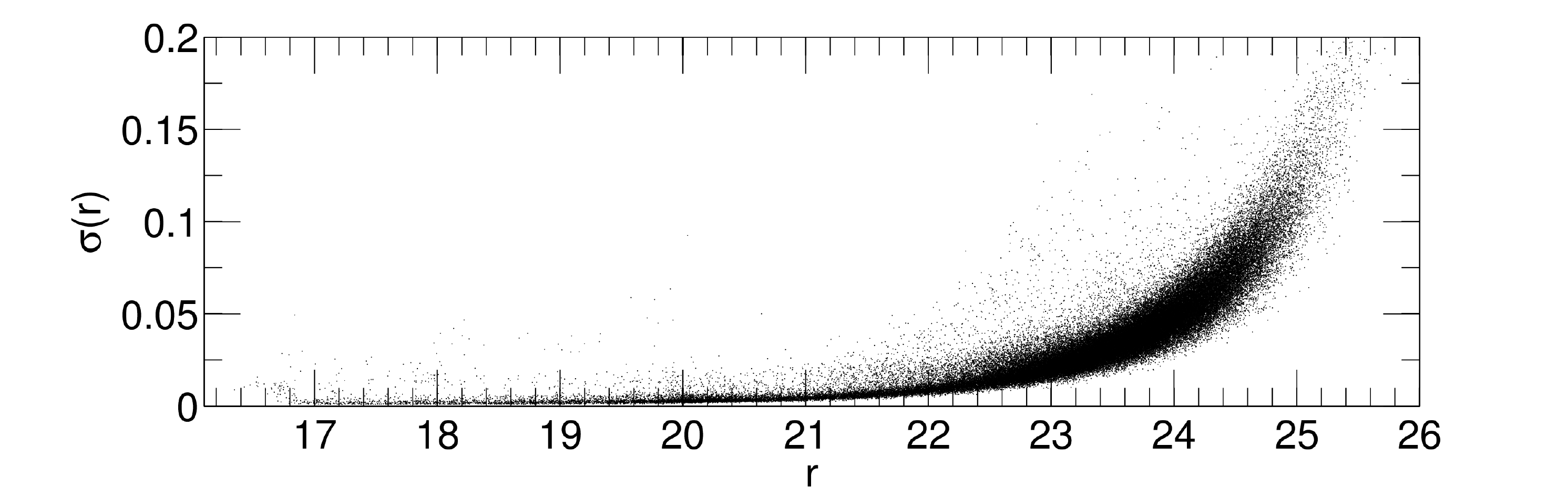}
   \caption[Calibrated magnitudes plotted against the corresponding photometric errors]{Calibrated magnitudes plotted against the corresponding photometric errors.}
\label{fig:errfoto}
\end{figure}

By means of a study of the chemical, kinematic and spatial distribution of its stellar populations on a region extended $40$ arcmin from the center of Sculptor, \citet{2004ApJ...617L.119T} discovered the presence of two distinct ancient stellar components (both $\geq 10$ Gyr). These two stellar components are characterized by different metallicities (one metal-rich $-0.9>[Fe/H]>-1,7$, one metal-poor $-1.7>[Fe/H]>-2.8$), distributions (the metal-rich is more centrally concentrated, while the metal-poor is spatially extended) and velocity dispersions ($\sigma_{metal-rich}=7 \pm 1$ km s$^{-1}$ and $\sigma_{metal-poor}=11 \pm 1$ km s$^{-1}$). 
\citet{2008ApJ...681L..13B} confirmed the above results ( on a radial distance of $\sim 1$ degree centered on the galaxy), moreover they measured a velocity gradient of $\sim 8$ km s$^{-1}$ deg$^{-1}$ along the projected major axis of Sculptor, likely due to intrinsic rotation. Also \citet{2005AJ....130.1065C} and \citet{2006AJ....131..375W} by means of wide-field medium resolution Ca II triplet spectroscopy of RGB stars independently confirmed the presence of two distinct components.
\citet{2016MNRAS.461L..41M} detected the presence of a significant metallicity spread ($\sim 2$ dex) within the RR Lyrae population, which is a population older than $10$ Gyr.
This suggests that Sculptor underwent an efficient early chemical enrichment fast enough to be recorded by the RR Lyrae.
\citet{2012A&A...539A.103D} presented the SFH of Sculptor based on deep wide-field B, V, I photometry taken with Mosaic II camera at the CTIO $4$ m Blanco telescope on an area of $1$ deg$^{2}$ centered on Sculptor \citep{2011A&A...528A.119D}. \citet{2012A&A...539A.103D} found that Sculptor is dominated by an old ($> 10$ Gyr) metal poor stars, but that younger, more metal-rich populations are also present.

In the present work, we take advantage of deep, wide-field ground-based photometry for deriving the SFH of the Sculptor dwarf spheroidal galaxy.
The article is organized as follows: in \S~\ref{sec:obs}, the observations and the data reduction are described along with the derivation of the photometry. The photometric calibration and the resulting colour-magnitude diagram (CMD) of Sculptor are presented too. In \S~\ref{sec:sfhh} we discuss completeness tests and error simulations. In this section is also illustrated the SFH derivation procedure. In \S~\ref{sec:discsculptor} the results are discussed. Finally, summary and conclusions are presented in \S~\ref{sec:concsculptor}.

\begin{table*}
\begin{minipage}{118mm}
\caption{The main properties of the Sculptor dSph}
\label{tab:param}
\begin{tabular}{@{}lcr}
\hline
Quantity & Value & References\footnote{(1) \citet{1991rc3..book.....D};  (2) \citet{2008AJ....135.1993P}; (3) \citet{1995A&A...300...31Q};  (4) \citet{1995MNRAS.277.1354I};  (5) \citet{2009MNRAS.394L.102L}.}\\
\hline
RA, $\alpha$ (J2000.0) &  $1^{\rm h} 00^{\rm m} 09^{\rm s}.4$ & (1)\\
Dec, $\delta$ (J2000.0) & -33$^{\circ}$ 42$^\prime$ 33.0$^{\prime\prime}$ & (1)\\
Galactic longitude, $l$ ($^\circ$) & 287.53 & (1)\\
Galactic latitude, $b$ ($^\circ$) & -83.16 & (1)\\
Galactocentric distance (kpc) & 86$\pm$5 & (2)\\ 
Heliocentric velocity (km s$^{-1}$) \hspace{20pt}& 109.9$\pm$1.4 & (3)\\
Ellipticity, $e$ & $0.32\pm 0.03$ & (4)\\
Position angle ($^\circ$) & 99$\pm$1 & (4) \\
Core radius ($^\prime$) & 5.8$\pm$1.6 & (4)\\
Tidal radius ($^\prime$) & 76.5$\pm$5.0 & (4)\\
Luminosity, $L_V$ ($L_\odot$) & ($2.03\pm0.79$)$\times 10^6$ & (5) \\
Absolute magnitude,$M_V$ & $-10.94\pm 0.58$ & (5)\\ 
Total mass, ($M_\odot$) & ($3.1\pm0.2$)$\times 10^7$ & (5) \\
Mass to light ratio, $M_\odot/L_\odot$ & $15.4\pm6.9$ & (5) \\
\hline
\end{tabular}
\end{minipage}
\end{table*}


\section{Observations and Data Reduction}\label{sec:obs}
Table~\ref{tab:obs} lists all the deep, wide-field exposures in $g$ and $r$ filters that have been used to construct the scientific stacked images. All the data were taken with DECam \citep{2015AJ....150..150F} at the Blanco 4m telescope at the Cerro Tololo Inter-American Observatory (CTIO), reduced with the NOAO Community Pipeline \citep{2014ASPC..485..379V} and retrieved from the NOAO Science Archive \citep{2002SPIE.4846..182S}. 
We obtained a stacked image for each filter as an average of all the scientific observations listed in Table~\ref{tab:obs} and we concentrated our analysis on the central CCDs centered on the galaxy, with a corresponding radial extension of $\sim 20$ arcmin from the center of Sculpor.
The total exposure time is $4800$ s in the $g$ filter and $4500$ s in the $r$ filter.


\begin{table*}
\centering
\caption{A summary of the available observational datasets}
\label{tab:obs}
\begin{tabular}{ccccccc} 
\hline
Calibration Images: & Filter & UT & Exposure(s) & Seeing('') & Program & PI\\
\hline
 & $g$ & $2013/08/19$ & $300\times1$ & $1.2''-1.4''$ & $2013$B-$0325[Scl]$ & Vivas\\
 & $g$ & $2013/08/19$ & $120\times9$  & $1.2''-1.4''$ & $2013$B-$0325[Stripe 82]$ & Vivas\\
 & $r$ & $2013/08/19$ & $300\times1$ & $0.9''-1.1''$ & $2013$B-$0325[Scl]$ & Vivas\\
 & $r$ & $2013/08/19$ & $120\times7$  & $0.9''-1.1''$ & $2013$B-$0325[Stripe 82]$ & Vivas\\
\hline
Scientific Images: & Filter & UT & Exposure(s) & Seeing('') & Program & PI\\
\hline
 & $g$ & $2013/08/19$ & $300\times16$ & $1.2''-1.4''$ & $2013$B-$0325$ & Vivas\\
 & $r$ & $2013/08/19$ & $300\times15$ & $0.9''-1.1''$ & $2013$B-$0325$ & Vivas\\
\hline
\end{tabular}
\end{table*}

The photometry on stacked images and on calibration images has been performed using the {\sc daophot/allstar} suite of programs \citep{1990ASPC....8..289S}. We used as point-spread function (PSF) a Moffat function with parameter $\beta=2.5$ and radius of $R_{PSF}=15$ pixels. A linearly varying PSF with the position of the stars has been derived for each stacked image by fitting $129$ stars in the $g$ filter and $60$ stars in the $r$ filter.
In order to reduce to 1\% the impact of the so called "brighter-fatter" effect \citep{2014JInst...9C3048A} we have chosen only stars with $5-10$ K adu peak, excluding in this way the brightest stars.
In order to remove poorly measured stars, and galaxies, only sources with $\sigma\leqslant0.2$ and $-0.5\leqslant$SHARP$\leqslant 0.5$ were considered from the output of {\sc allstar}.
In Fig.~\ref{fig:errfoto} are plotted the magnitudes and the corresponding photometric errors for each filter with the restrictions above applied.

With the two catalogues in the $g$ and $r$ filters in hand, we performed the match using the packages {\sc daomatch} and {\sc daomaster} \citep{1993spct.conf..291S}.
The final total catalogue, resulting from the match above, in $g$ and $r$ filters contains $\sim 96000$ stars.
We performed the aperture corrections for each filter, from the comparison between aperture and PSF magnitudes, finding for the $g$ filter $0.0121$ mag and for the $r$ filter $0.0351$ mag.
Then we corrected for the exposure time, in $g$ filter applying $+2.5\times\log(300)$, while in $r$ filter, $+2.5\times\log(300)$. The quantities into the parenthesis have been calculated as the average of each exposure time in each filter.


In order to calibrate the photometry derived from the stacked images for each filter we have performed a two step calibration. The final scientific stacked images are composed among the others by a set of observations performed by Vivas (see last four rows in Table~\ref{tab:obs}).
Thus, we firstly performed PSF photometry on the stacked images of Sculptor by Vivas, obtaining the catalog in $g$ and $r$ filters. 
Then, we retrieved the PSF photometry of the single images of the standard stars observed the same night, see Stripe 82 observations in Table~\ref{tab:obs}. We indentified the standard stars by using the SDSS Stripe $82$ Standard Stars Catalog \citep{2007AJ....134..973I}. The magnitudes of photometric standard stars has been compared to the SDSS Stripe $82$ catalog in order to obtain the extinction coefficients for that night for both filters. Knowing the extinction coeffiecients and the photometry of standard stars, we derived the equations for the standard calibration:

\begin{align}
\label{eqn:calib1}
\begin{split}
g'-g &= k_{g}\times X_{g}+c_{g1}\times(g'-r')+g_{0}
\\
r'-r &= k_{r}\times X_{r}+c_{r1}\times(g'-r')+r_{0}
\end{split}
\end{align}

where $g'$, $r'$ are the calibrated magnitudes, $g$, $r$, the observed magnitudes, $k_{g}$, $k_{r}$ the airmass coeffiecients, $X_{g}$, $X_{r}$ the effective airmasses, $c_{g}$, $c_{r}$, the colour therms and $g_{0}$, $r_{0}$, the photometric zero points. 
We derived the photometric solution using a polynomial regression model for the three independent variables.
The root mean square error associated to this transformation are $0.0158$ mag in $g$ filter and $0.0201$ mag in $r$ filter, hence indicating a good match with the standard system. 

Using the system of equations above we applied the calibration to the Vivas' catalog. In this way these stars can serve as local standards.
Then, from the comparison between the magnitudes of the local standards with the full catalog, we obtained the equations to calibrate the full catalog:

\begin{align}
\label{eqn:calib2}
\begin{split}
g'-g &= c_{g2}\times(g'-r')+z_{g}
\\
r'-r &= c_{r2}\times(g'-r')+z_{r}
\end{split}
\end{align}

where $c_{g2}$, $c_{r2}$ are the colour therms and $z_{g}$, $z_{r}$, the photometric zero points. 
In Table~\ref{tab:calib} are listed all the coeffiecients relative to the two obtained photometric tranformations, with the corresponding errors. 


\begin{table}
\centering
\caption{Airmass coefficients, colour therms and photometric zero points adopted for the calibration. The associated errors are also listed.}
\label{tab:calib}
\begin{tabular}{cc} 
\hline
Parameter & Values\\
\hline
$k_{g}$ &  $-0.1514\pm0.0064$\\
$c_{g1}$ &  $0.1157\pm0.0029$\\
$g_{0}$ &  $25.4106\pm0.0083$\\
\hline
$k_{r}$ & $-0.1264\pm0.0082$\\
$c_{r1}$ & $0.0918\pm0.0037$\\
$r_{0}$ & $25.5771\pm0.0106$\\
\hline
$z_{g}$ & $25.3062\pm0.0008$\\
$c_{g2}$ & $0.1113\pm0.0014$\\
\hline
$z_{r}$ & $25.4677\pm0.0007$\\
$c_{r2}$ & $0.0902\pm0.0012$\\
\hline
\end{tabular}
\end{table}

\subsection{The Colour-Magnitude Diagram}
\begin{figure*}
\includegraphics[width=15 cm]{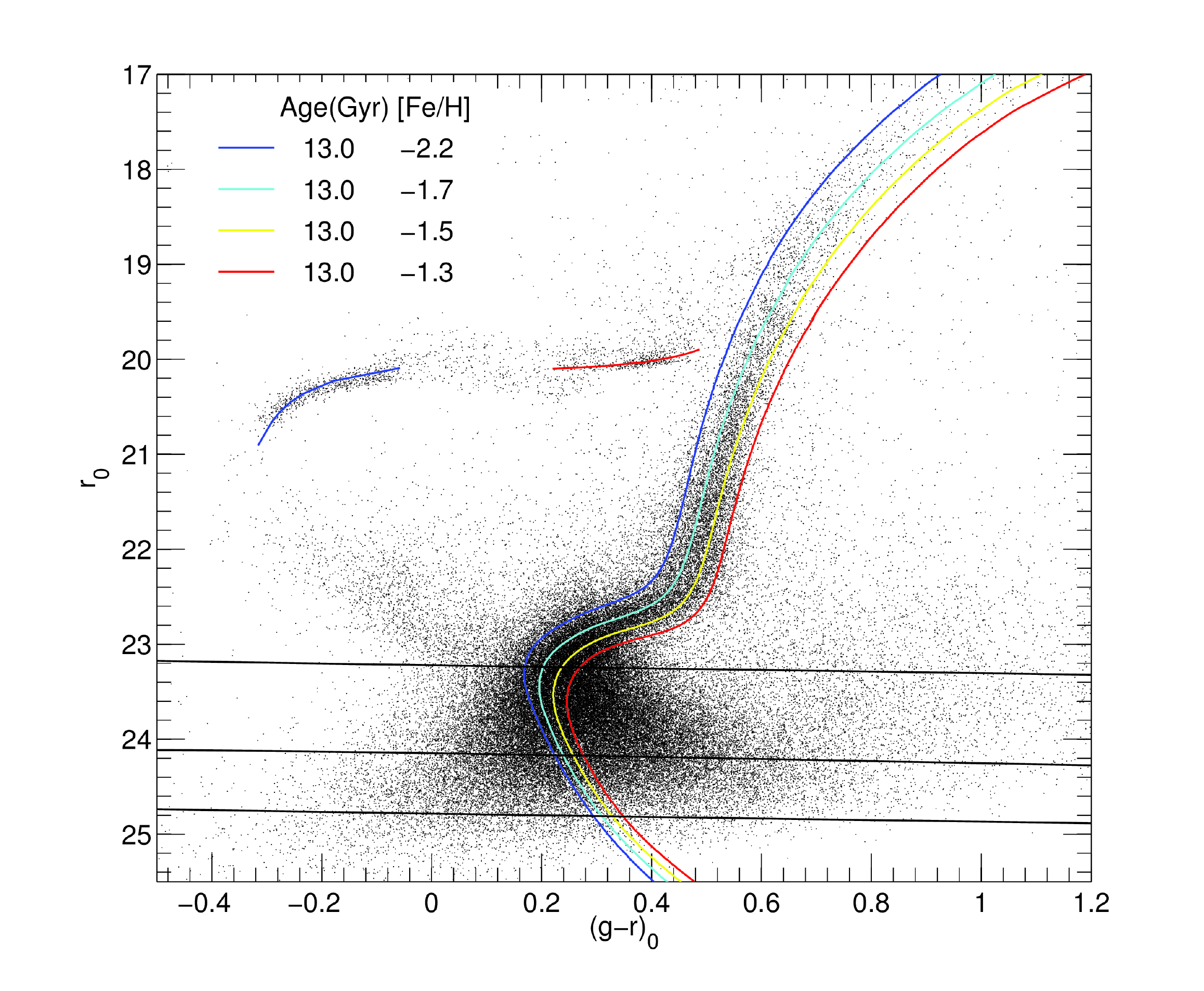}
   \caption[Calibrated CMD of Sculptor]{Observed CMD of Sculptor dSph. Four BaSTI isochrones have been superimposed on the CMD, see the labels for details.
   The Blue Horizontal Branch (BHB) and Red Horizontal Branch (RHB) are fitted with a Zero Age Horizontal Branch (ZAHB) of $[Fe/H]=-2.2$ and $[Fe/H]=-1.3$ respectively (solid blue and red lines in the electronic version).
   Completeness levels are over-plotted as black lines for values corresponding to $90\%$, $75\%$ and $50\%$}
\label{fig:cmd0}
\end{figure*}

The resulting CMD of Sculptor extends $\sim 2$ mag below the MS turn-off (see Figure \ref{fig:cmd0}), thus allowing us to extract the information relative to star formation also for the earliest epochs.
Figure \ref{fig:cmd0} shows the resulting CMD, corrected by reddening ($A_{g}=0.06$, $A_{r}=0.042$ (\citealt{1998ApJ...500..525S}; \citealt{2011ApJ...737..103S})), with over-plotted four BaSTI isochrones \citep{2004ApJ...612..168P} adopting as distance modulus $(m-M)_{0}=19.57$ \citep{2011MNRAS.414.3492M}. 
The main-sequence (MS) is well matched by isochrones of $[Fe/H]=-1.7$ and $[Fe/H]=-1.5$ and an age of about $13$ Gyr.
The RGB locus is very broad showing a spread in $[Fe/H]$ between $-2.2$ and $-1.3$. 
Actually, an exhaustive spectroscopic analysis of the RGB has shown that this dwarf seems to have a significant spread that varies with position on the RGB \citep{2011A&A...528A.119D}.
In the obtained CMD there is a region ($-0.3\lesssim(g-r)_{0}\lesssim0.08$, $21.5\lesssim r_{0} \lesssim23.2$) that appears to be populated by Blue Straggler stars (BSS) (\citealt{2009MNRAS.396.1771M}; \citealt{2012ApJ...744..157M}).

\section{Derivation of the SFH}\label{sec:sfhh}
The SFH of Sculptor has been derived following the prescriptions of \citet{2009AJ....138..558A} and \citet{2011ApJ...730...14H}.
The star formation rate as a function of time and the age-metallicity relation have thus been computed.
We made use of IAC-Star \citep{2004AJ....128.1465A} for the synthetic CMDs (sCMDs) computation. IAC-pop \citep{2009AJ....138..558A}, which is the key algorithm for solving for the best parametrization of the SFH. Finally, for sampling the parameter space, creating input data to IAC-pop and averaging solutions, MinnIAC \citep{2011ApJ...730...14H} was used. 

\begin{figure}
\includegraphics[width=\columnwidth]{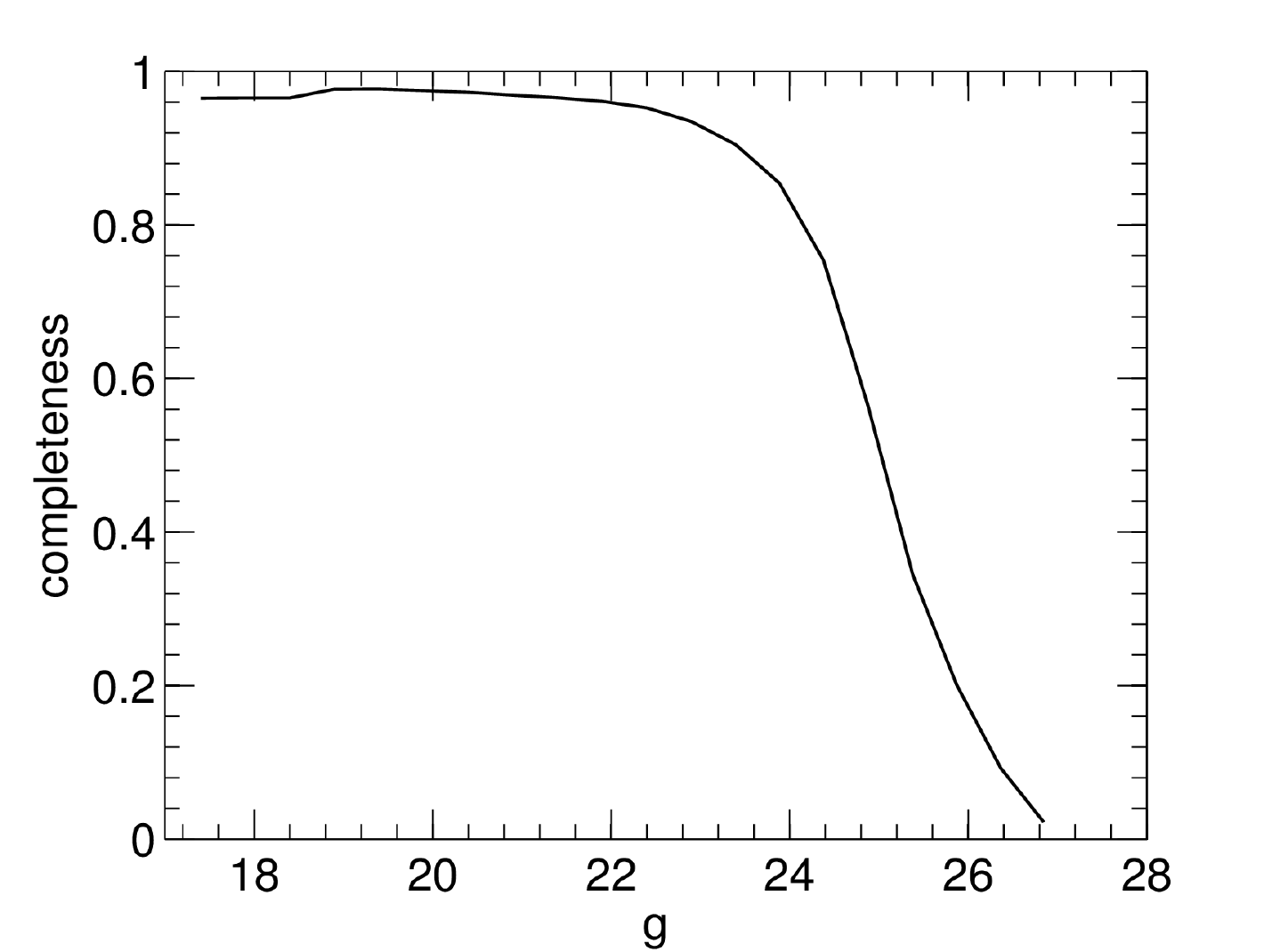}\\
\includegraphics[width=\columnwidth]{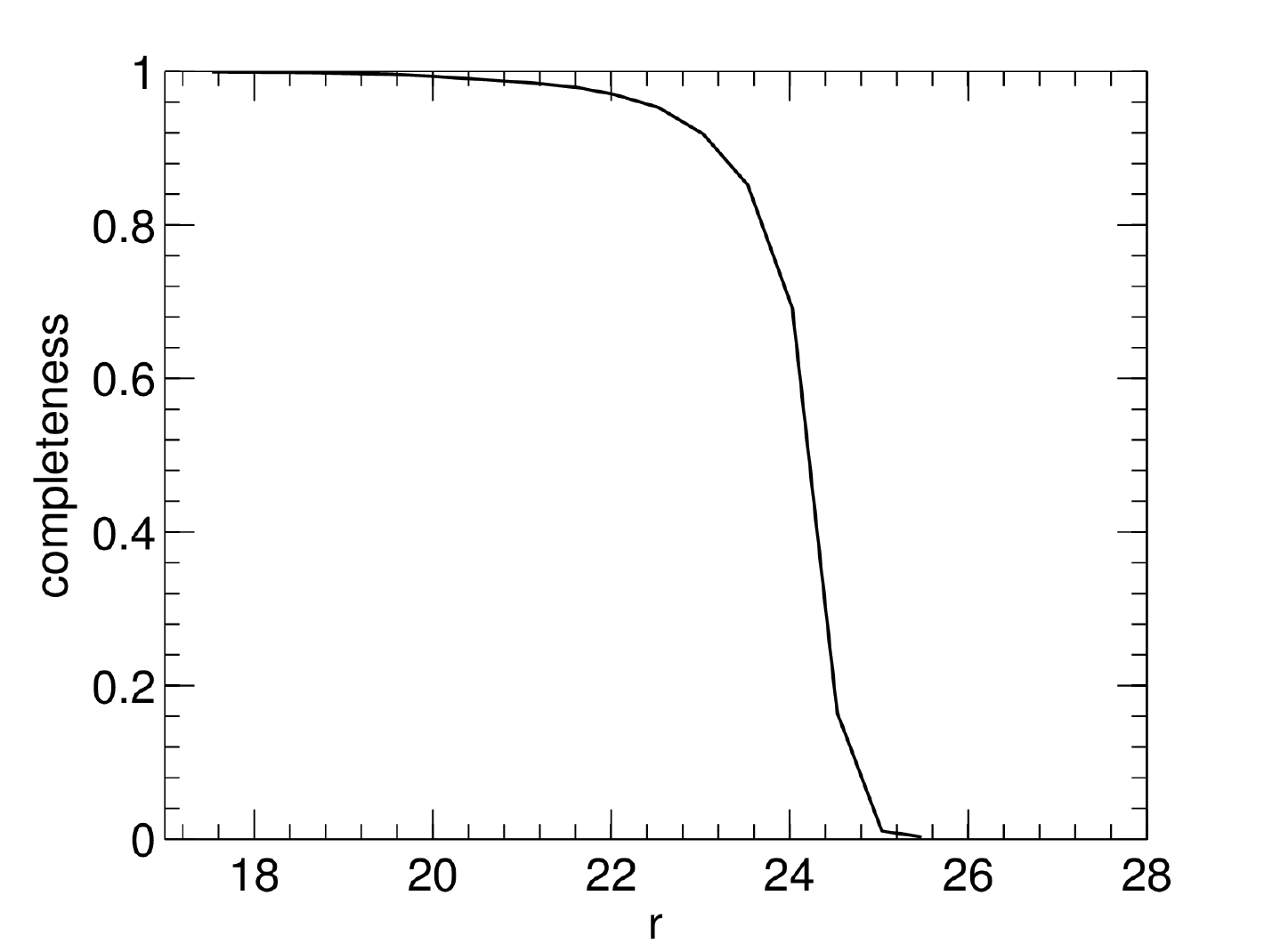}
   \caption[]{Results of the completeness tests as a function of $g$ (upper panel) and $r$ (lower panel) filter in $0.5$ mag bins.}
\label{fig:comp}
\end{figure}

Using {\sc IAC}-star with the BaSTI stellar evolutionary library, we created a synthetic CMD (sCMD), which is used to derive the stellar properties of Sculptor. It is composed by $5\times10^{6}$ stars, with a constant star formation rate (SFR) from $0$ to $13.5$ Gyr and a flat metallicity distribution in the interval $0.0001\leqslant$ Z $\leqslant 0.002$ for all ages. The \citet{2011ApJ...727...78K} metallicity distribution function (MDF) has been used as a template to determine this metallicity interval. 
We adopted the bolometric corrections for the SLOAN photometric systems provided by  \citet{2004ApJ...612..168P}, based on the model atmospheres by \citet{1993ASPC...44..496C}.
A \citet{2002Sci...295...82K} initial mass function (IMF) has been adopted and a population of binary stars has also been simulated. For this case we assumed $f=0.3$ for the fraction of binary stars and $q_{min}=0.5$ as the minimum secondary to primary stellar mass ratio $q$ with a flat distribution for this last parameter.
An accurate modelling of observational effects is a crucial step in the obtainment of a realistic model CMD to be compared with the real data \citep{1995AJ....110.2105A}.
As outlined in \citet{2011ApJ...730...14H}, $5\times10^{6}$ artificial stars has been injected in each stacked image along an uniform grid, with a separation of at least $2 \times R_{PSF}+1$ pixels between the centroids of each of them. Then, the photometry has been performed again on these stacked images, using the same configuration as done for the original ones. 
Magnitude values of the injected artificial stars have been generated according to $-1.25\leqslant (g-r) \leqslant 2$ and $16.75 \leqslant g,r \leqslant 26$ in order to cover the entire range of luminosity and colour of the observed CMD (oCMD) (see Fig.~\ref{fig:cmd0}). 
Since the the turn-off (TO) region ($22 \leqslant g,r \leqslant 26$) is of highest interest for the derivation of the SFH, $3.5\times10^{6}$ stars have been injected in this magnitude range, while, $1\times10^{6}$, have been generated in the red-giant-branch (RGB) region ($19 \leqslant g,r \leqslant 22$). Finally, $0.5\times10^{6}$ artificial stars have been injected in the upper part of the colour-magnitude diagram ($16.75 \leqslant g,r \leqslant 19$). The results of the completeness tests for each filter are shown in Figure \ref{fig:comp}.

\begin{figure}

\includegraphics[width=\columnwidth]{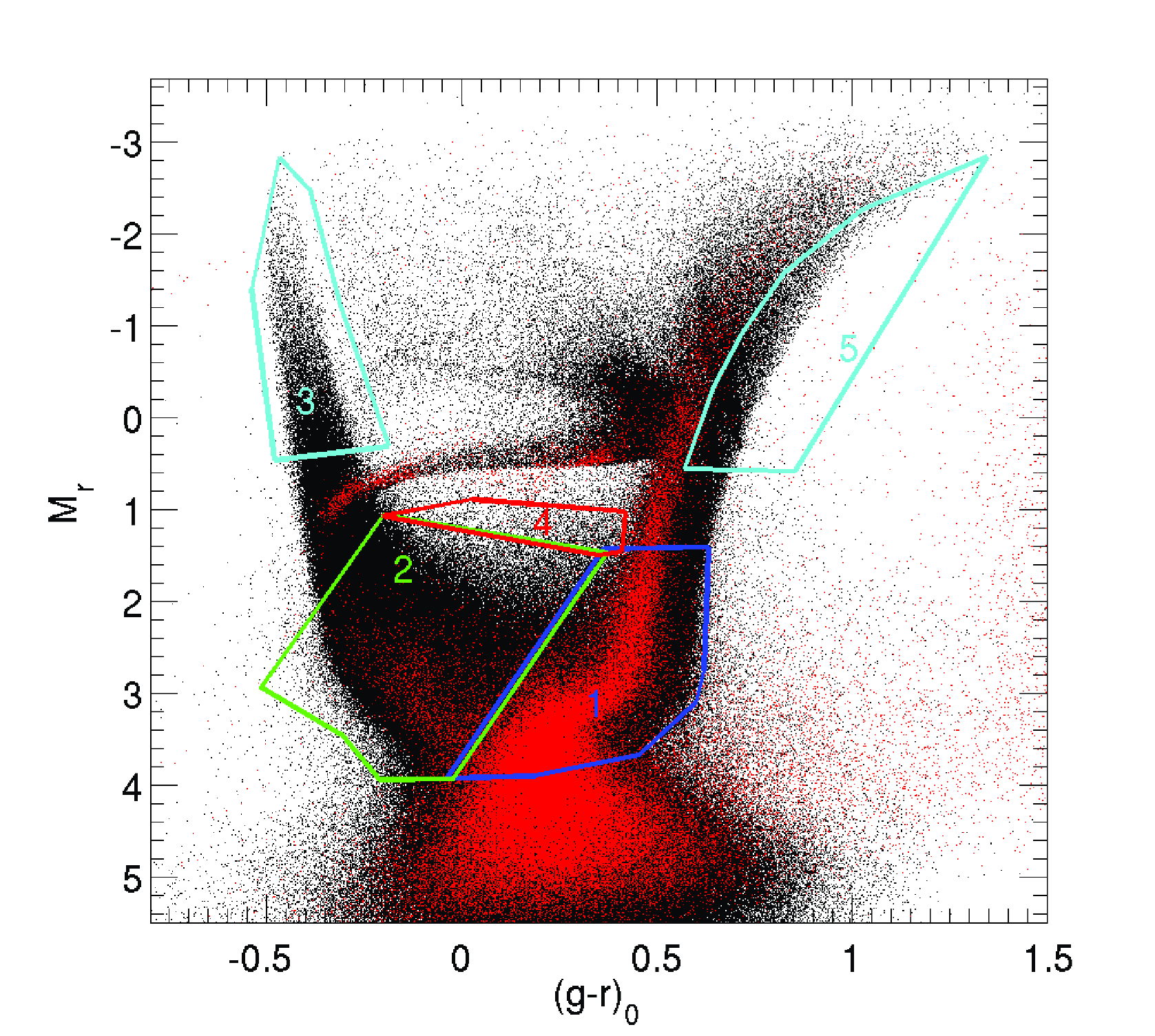}
   \caption[Synthetic CMD with observational effects simulated with over-plotted the observed CMD of Sculptor]{ Synthetic CMD with the simulated observational effects (black). The observed CMD is also shown (red). Regions numbered from  $1$ to $5$ are the adopted \textit{bundles}, see text for details.
}
\label{fig:simul}
\end{figure}

In order to measure the completeness level, the ratio between the number of the recovered artificial stars and the totality of injected ones in each colour and magnitude interval has been calculated. In Fig.~\ref{fig:cmd0}, the $50\%$, $75\%$ and $90\%$ completeness levels are over-plotted on the CMD of Sculpor.
Moreover, the position on the image of each artificial star, along with the injected magnitude $m_{inj}$ and the recovered one $m_{rec}$, have been recorded in order to simulate the observational effects, see \citet{2011ApJ...730...14H} for a detailed description of the procedure.
Figure~\ref{fig:simul} shows the sCMD accounting for the observational effects simulation.


To obtain the SFH, the sCMD has been subdivided into simple stellar populations (SSPs) and the resulting star distributions have been compared with that of the oCMD. 
The sampling of the sCMD into SSPs has been performed binning it in age and metallicity. For the age, intervals of $2$ Gyr for the first $8$ Gyr have been chosen and a smaller one, of $0.5$ Gyr, for the last interval between $8$ Gyr and $13.5$ Gyr, since higher resolution is necessary to well characterize the earliest SFH phase.  
For the metallicity, we have adopted the following bins: $(0.01,\,\,\, 0.1,\,\,\,  0.3,\,\,\,  0.5,\,\,\,  1,\,\,\,  2) \times 10^{-3}$. Wider intervals have been chosen for higher $Z$ values.

To sample the oCMD and the sCMD, \textit{bundles} have been defined, i.e. macroregions which identify main features of the CMDs, as shown in Fig.~\ref{fig:simul}.

Each \textit{bundle} has a weight on the solution, given by the number of boxes of fixed size in color and in magnitude defined in them. More boxes are present in a \textit{bundle}, the larger will be its weight on the resulting SFH. Table~\ref{tab:boxbundle} shows the weight of each bundle in the solution.

The dependence of the resulting SFH on the CMDs sampling parameters has been taken into account obtaining $24$ solutions varying the CMD binning within each bundle and the SSPs sampling.
In particular, we have shifted by a $30\%$ the age and metallicity in a total of $12$ combinations of SSPs, each one sampled with two slightly different combinations of boxes distribution in the bundles. Moreover, the effect of uncertainties in the distance modulus, reddening and photometric calibration, has been mitigated shifting the oCMD $25$ times along a regular grid with nodes in colour $\Delta(g-r)=[-0.1,-0.05,\,\, 0,\,\, 0.05,\,\, 0.1]$ and magnitude $\Delta r=[-0.2, -0.1,\,\, 0,\,\, 0.1,\,\, 0.2]$. 
For each node $24$ solutions have been calculated as described above, obtaining a total of $600$ solutions. A mean solution $\overline{\psi}$ and its $\overline{\chi^{2}_{\nu}}$ have been also calculated for each node. The $\overline{\psi}$ have been obtained by using a boxcar of width $0.5$ Gyr and step $0.1$ Gyr for $t$ and width $0.00025$ and step $0.00005$ for $Z$. In this way $25$ different $\overline{\chi^{2}_{\nu}}$ have been obtained, the best solution was then selected as the one with the lowest value.
In this particular case, the best solution is the one corresponding to a shift of $-0.05$ mag in colour and no shift in $r$ magnitude.

For an extensive description of the five bundles adopted and how we performed the minimization of the solution on the CMDs sampling parameters we refer to \citet{2011ApJ...730...14H} (see also \citet{2018MNRAS.476...71B}).


\begin{table}
\centering
\caption{Box sizes in each bundle that sample the observed CMD}
\label{tab:boxbundle}
\begin{tabular}{cccc} 
\hline
Bundle \# & $\Delta col$ & $\Delta mag$ & $Boxes$\\
\hline
$1$ & $0.025$ & $0.125$ & $1020$\\
$2$ & $0.1$ & $0.205$ & $150$\\
$3$ & $0.1$ & $0.23$ & $30$\\
$4$ & $0.1$ & $0.23$ & $3$\\
$5$ & $1.5$ & $0.67$ & $1$\\
\hline
\end{tabular}
\end{table}

\subsection{The Global SFH of Sculptor}\label{sec:casesculptor}
\begin{figure}

\includegraphics[width=\columnwidth]{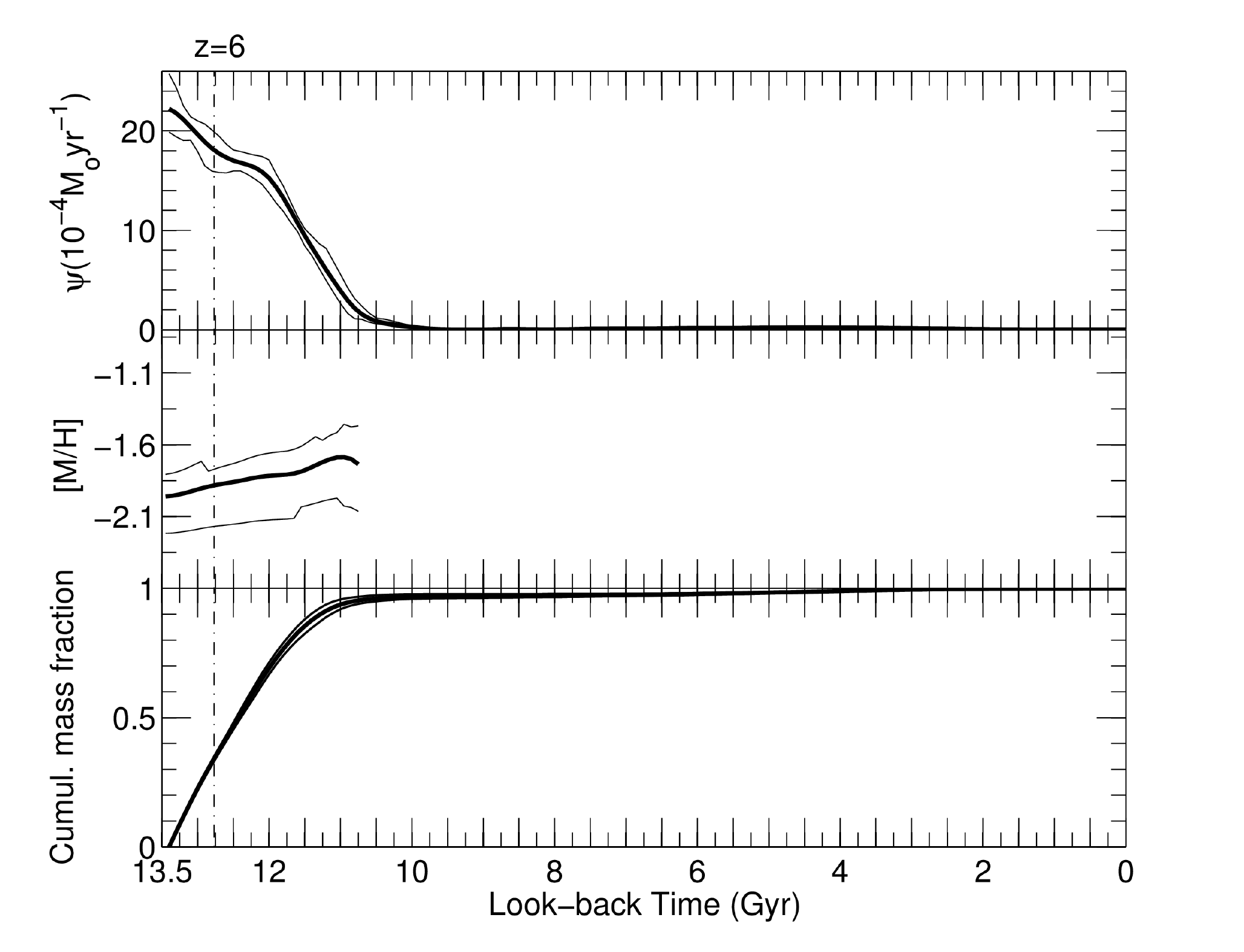}
   \caption[Total SFH for Sculptor]{Results of the Sculptor SFH. Top panel: SFH as a function of time ($\psi(t)$). Middle panel: metallicity of the system as a function of the time. Lower panel: cumulative mass fraction as a function of the time. $1 \sigma$ uncertainties have been drawn as thin lines.}
\label{fig:totsfh}
\end{figure}

\begin{figure}

\includegraphics[width=\columnwidth]{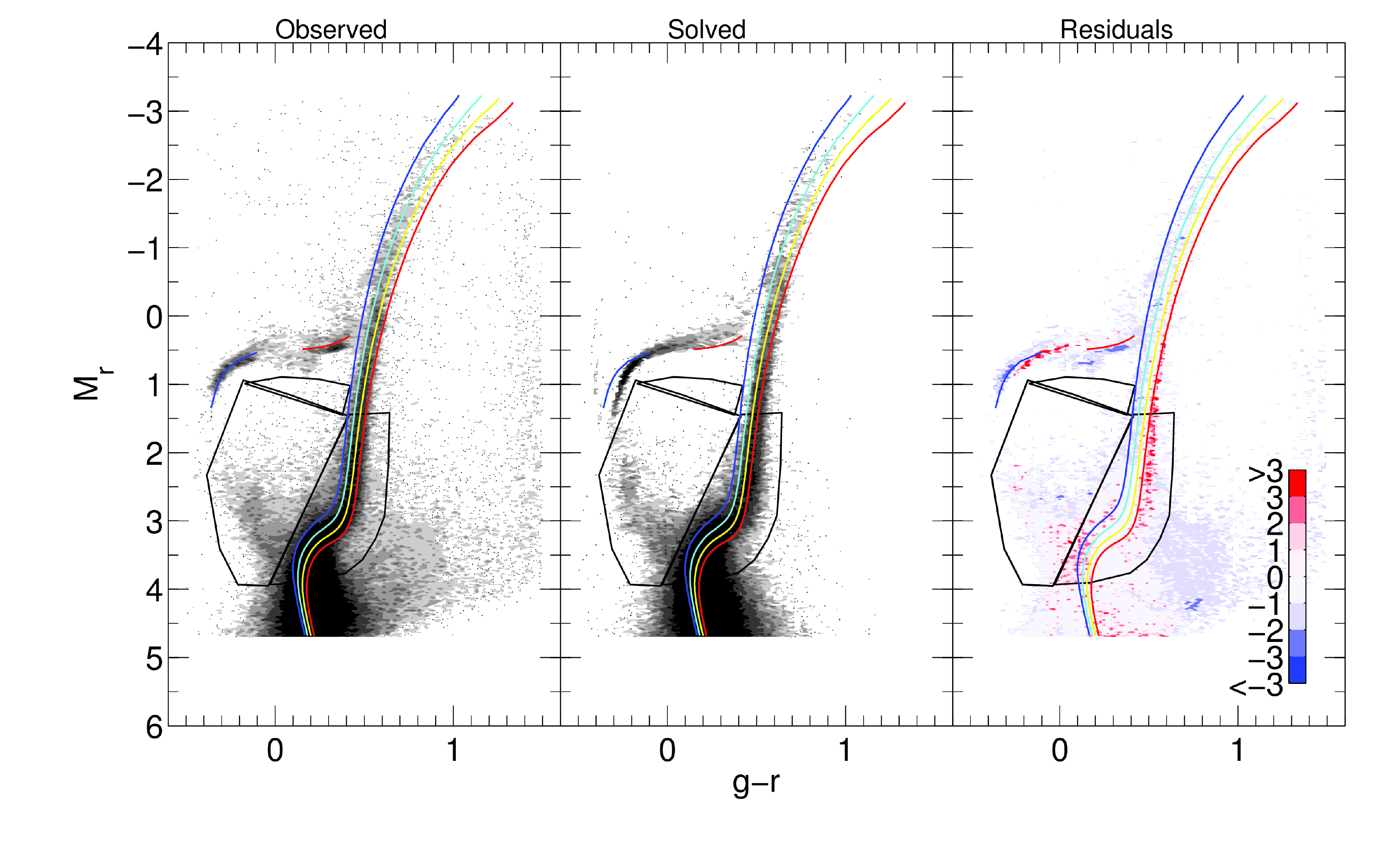}
   \caption{Left panel: oCMD Hess diagram. Gray levels are proportional to the density of stars. A factor of $2$ in density exists between each two successive gray levels. The single dots are shown where the density is less than $2$ stars per $(0.02)^2$ mag. Middle panel: best solution CMD Hess diagram.  Right panel: residuals CMD Hess diagram in units of Poisson uncertainties. Values of $\sigma<0$ are found when the oCMD contains more star than the best solution CMD. The same isochrones as in Figure~\ref{fig:cmd0} are over-plotted. The most relevant \textit{bundles} used are also shown.}
\label{fig:residuototale}
\end{figure}

Fig.~\ref{fig:totsfh} shows the star formation rate as a function of time, $\psi(t)$, the age-metallicity relation, $Z(t)$, and the cumulative mass function for Sculptor.
The age resolution of our derived SFH indicates that Sculptor has experienced a single event of star formation limited to the first $\sim 2$ Gyrs after Big Bang, producing $\sim 70 \%$ of its mass about $12$ Gyr ago.
The mean metallicity retrieved is $[Fe/H]\sim \,-1.8$.
In Fig.~\ref{fig:residuototale} the Hess diagrams of the oCMD (left panel) and the best solution CMD (middle panel) are shown. The residuals diagram, with values expressed in units of Poisson error, is also shown in the right panel. It can be noted that there is good agreement between oCMD and the best solution CMD in the metallicity range $-2.2<[Fe/H]<-1.3$, where the bulk of the distribution of stars lies.
Out of this range, the residuals progressively worsen indicating a relative excess of model with respect to the observed distribution. For a stellar population of age fixed at $13$ Gyr old, the bad residuals in the blue region of the MS would show an overestimation of the more metal-poor stars, while the bad residuals in the red side of the RGB would be produced by an overestimation of the metal-rich stars. 
This effect could be due, at least in part, to some still-present shortcomings in the available color-$T_{eff}$ relations used for transferring the stellar models from the theoretical plane to the observational ones.

\subsection{The Radial SFH of Sculptor}\label{sec:radialsculptor}
\begin{figure}
\includegraphics[width=\columnwidth]{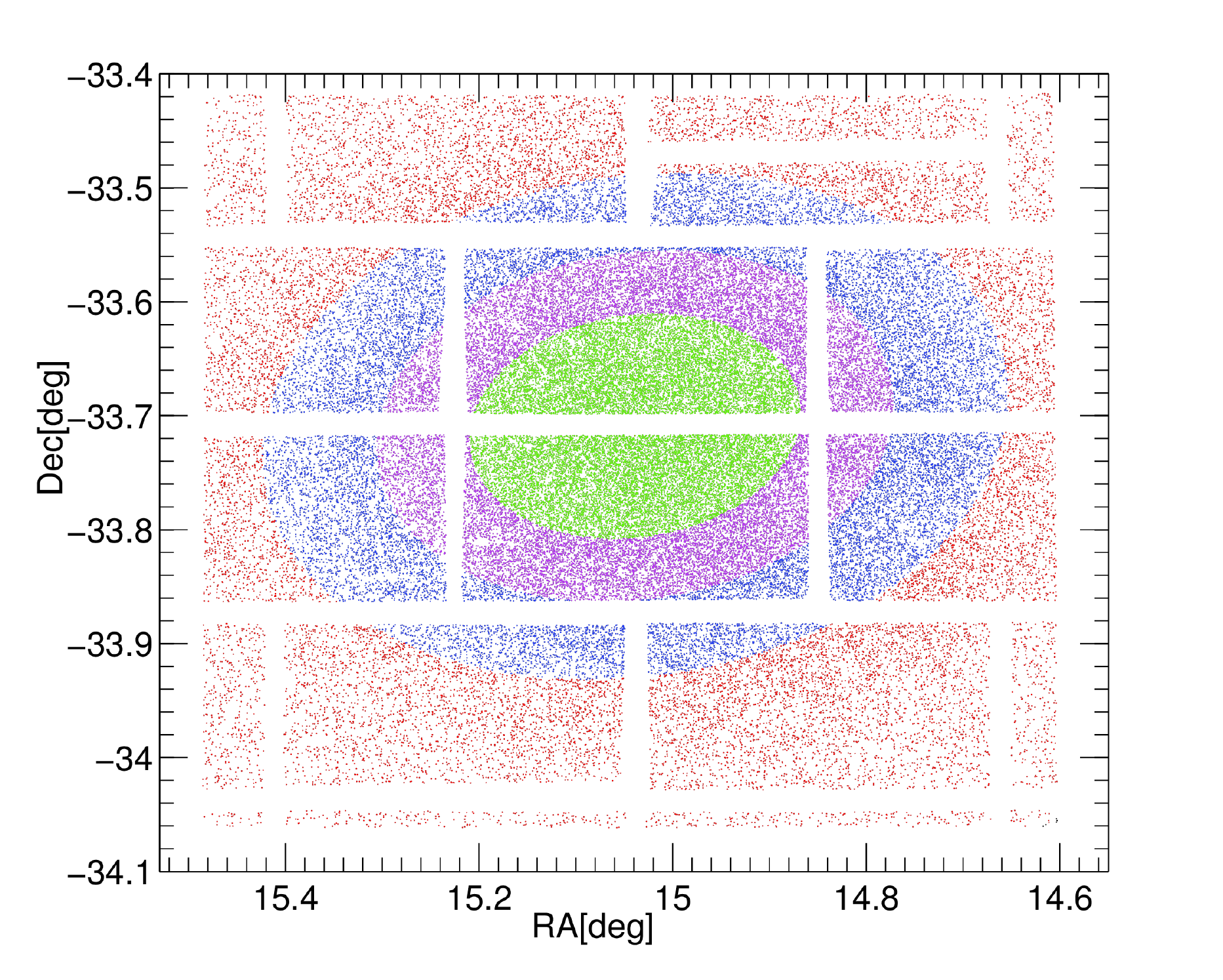}
   \caption[Stellar spatial distribution]{Stellar spatial distribution. The ellipse major axis are in the intervals $a\leq8.59$ arcmin for green points; $8.59 <a\leq 13.41$ arcmin for magenta points; $13.41 <a\leq 19.35$ arcmin for blue points; and $a\geq 19.35$ for red points. The region of the core radius is well represented by the innermost elliptical region (green points).
}
\label{fig:circless}
\end{figure}

\begin{figure*}
\includegraphics[width=15cm]{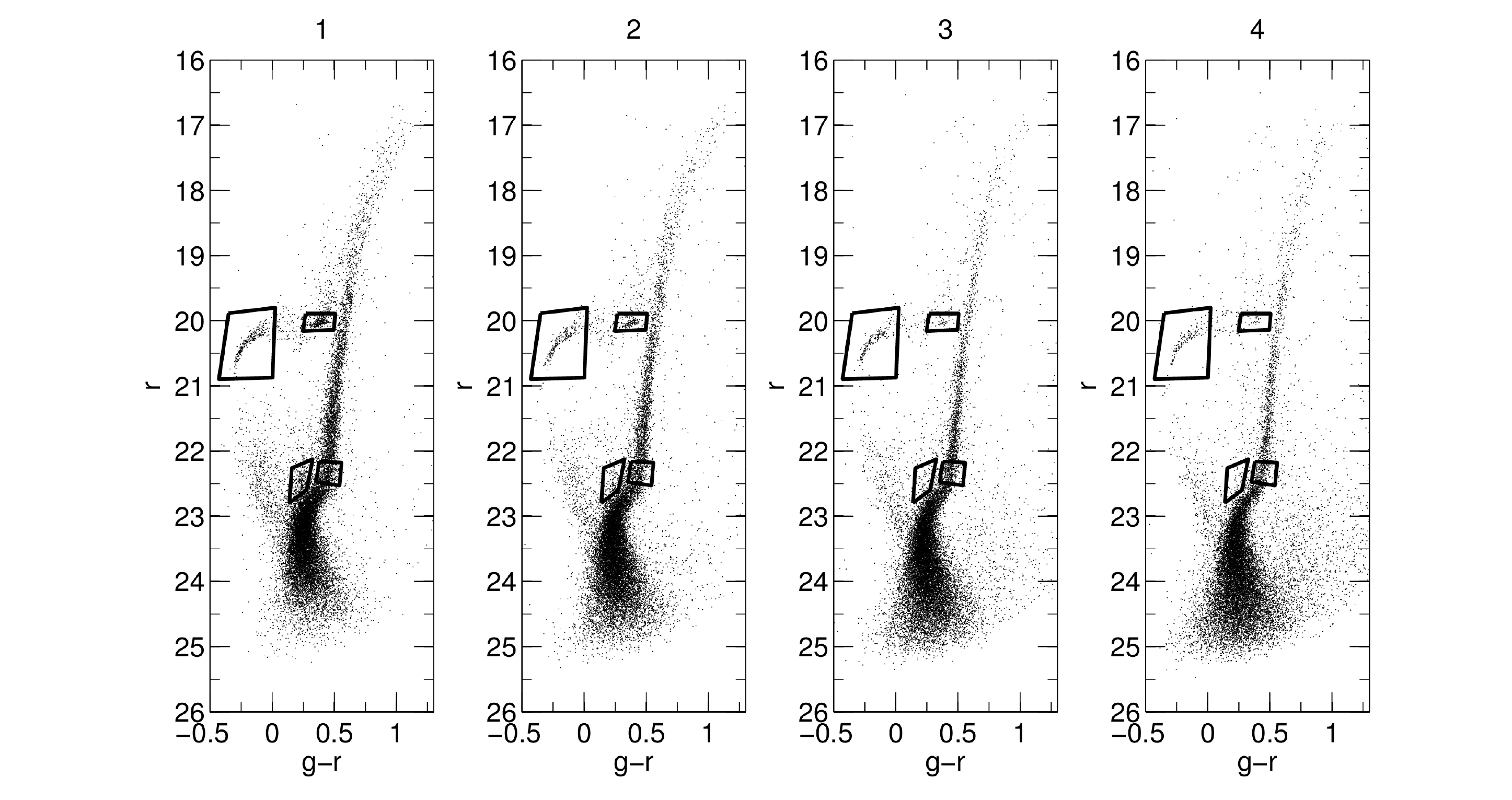}
   \caption[From left to right CMDs of Sculptor from the innermost elliptical region to the outer]{From left to right CMDs of Sculptor from the innermost elliptical region to the outer. Region $1$ refers to the elliptical area within the major axis $a\leq 8.60$ arcmin, region $2$ to the major axis $8.60\leq a\leq 13.42$ arcmin, region $3$ to the major axis $13.42\leq a\leq 19.35$ arcmin and region $4$ to the major axis $19.35\leq a\leq 39.46$ arcmin. On the CMDs are drawn boxes indicating the region of the BHB, RHB and the spur region, near the MS. A box is indicated at the base of the RGB, see text for details.}
\label{fig:hk}
\end{figure*}

To investigate whether a radial gradient of stellar populations exists in Sculptor, we derived the SFH by dividing the sampled area into four elliptical regions with delimiting major axis of $8.59$, $13.41$, $19.35$ and $39.46$ arcmin, see Fig.~\ref{fig:circless}. The center assumed for the analysis is coincident with the one tabulated by \citet{1938BHarO.908....1S} in J$2000.0$ coordinates $(15.03898,-33.70903)$. The position angle adopted is $\theta=99$ and ellipticity $0.32$  \citep{1995MNRAS.277.1354I}.
The major axes have been fixed in order to have $24000$ stars in each elliptical region for statistical consistency. Figure~\ref{fig:hk} shows the CMDs for each elliptical region in which the galaxy has been divided.


For each region the SFH has been derived in the same ways for the global one but using in each case the corresponding oCMD (see Fig.~\ref{fig:hk}). 

Regarding the morphological changes connected to stellar populations variations, it is remarkable  how the horizontal branch (HB) changes with distance: a populated red HB (RHB) is present in the center but it seems to be less populated moving outwards, while blue HB (BHB) stars seem constant along all the radial distance. This suggests the presence of an extended old metal poor stellar population and a more centrally concentrated old more metal rich stellar population.

In order to obtain a deeper insight on this effect, we calculated the ratios between the number of stars in the RHB, BHB and the so called \textit{spur} \citep{1999ApJ...523L..25H} region in respect to the RGB region indicated in Fig.~\ref{fig:hk}. In Table~\ref{tab:ratios} we report these ratios. They indicate that effectively RHB stars are more centrally concentrated while BHB stars more extended.
This fact has been already shown in \citet{2004ApJ...617L.119T}.
The RGB seems to narrow when moving from region $1$ to region $4$, indicating a clear change in stellar populations.
In Figure \ref{fig:hk}, we  also indicate a box that extends from the MS TO region and is located between the canonical blue stragglers (BS) region and the SGB, which englobes the \textit{spur}, a feature firstly identified in \citep{1999ApJ...523L..25H}.
The same feature is visible in the data presented in this work and it is interesting to note that this feature becomes weaker with the distance from the center of the galaxy. In particular, from the calculated ratios it appears that also the \textit{spur} is more centrally concentrated, but at a lower extent, when compared to the RHB.

\begin{table}
\centering
\footnotesize
\caption{Ratios between the number of stars in the RHB, BHB and the spur region in respect to the RGB}
\label{tab:ratios}
\begin{tabular}{cccc} 
\hline
Region & $RHB/RGB$ & $BHB/RGB$ & $spur/RGB$\\
\hline
$1$ & $0.51$ & $0.50$ & $0.55$\\
$2$ & $0.32$ & $0.47$ & $0.38$\\
$3$ & $0.16$ & $0.56$ & $0.27$\\
$4$ & $0.11$ & $0.52$ & $0.25$\\
\hline
\end{tabular}
\end{table}

\begin{figure}

\includegraphics[width=\columnwidth]{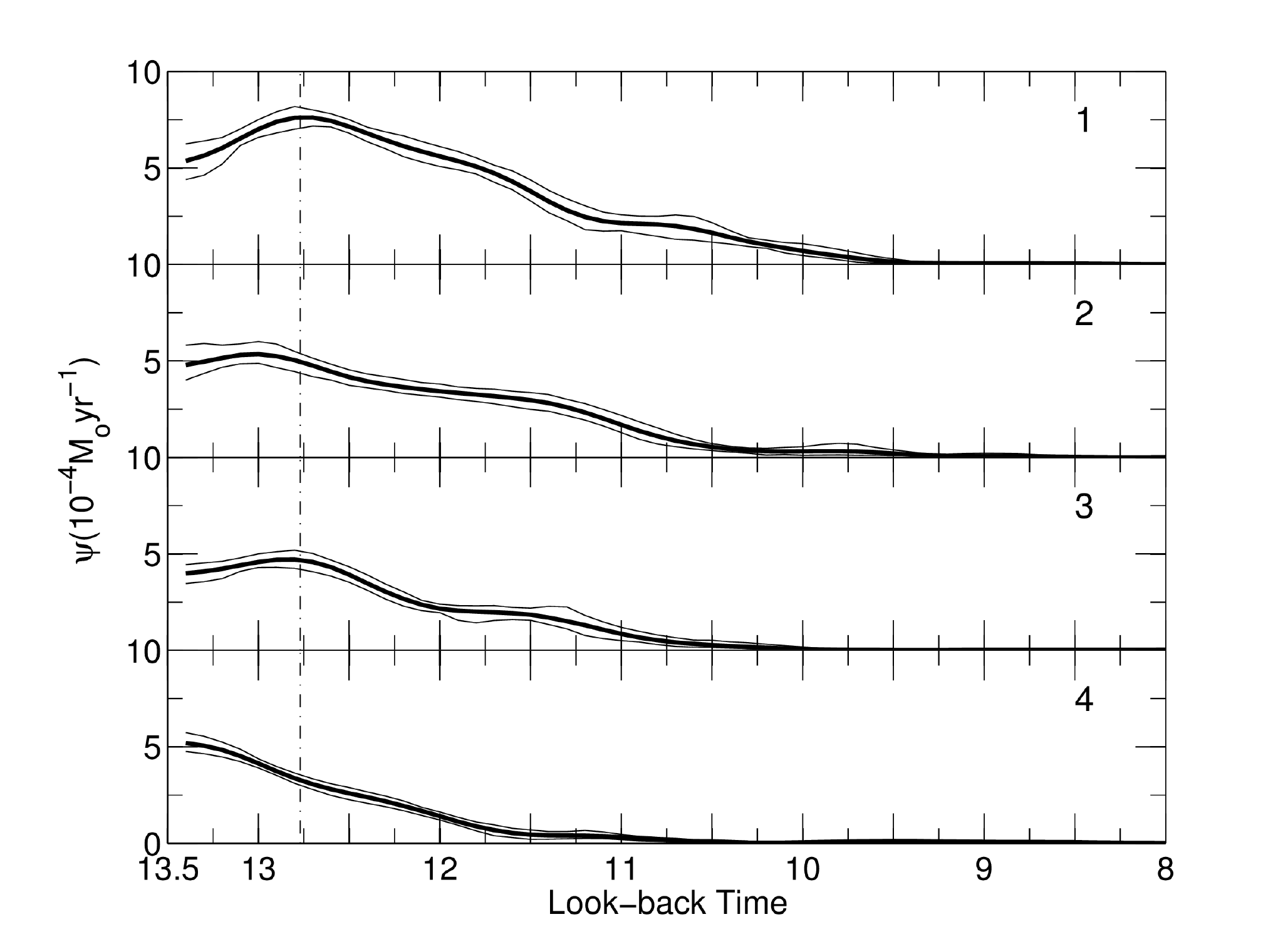}\\
\includegraphics[width=\columnwidth]{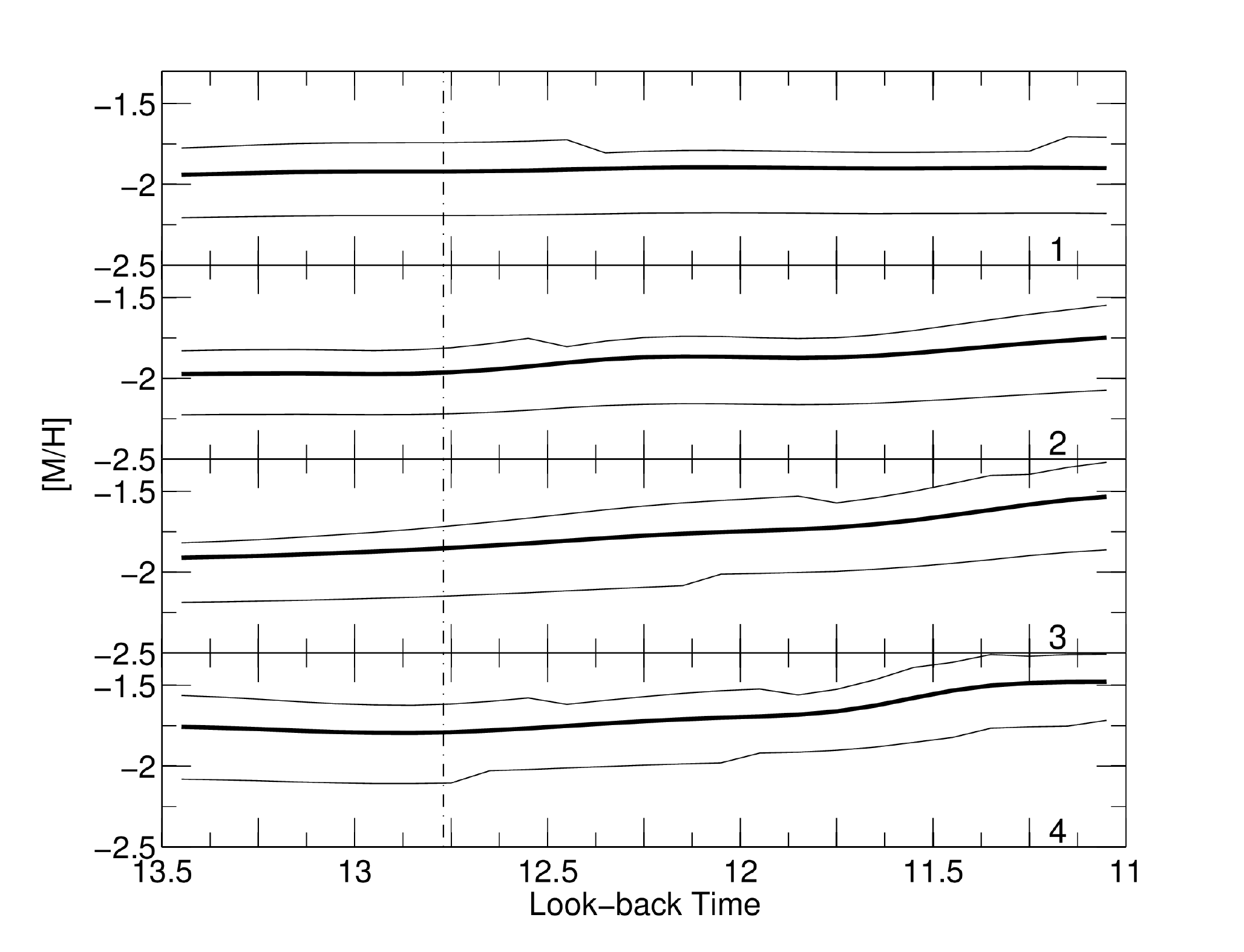}\\
\includegraphics[width=\columnwidth]{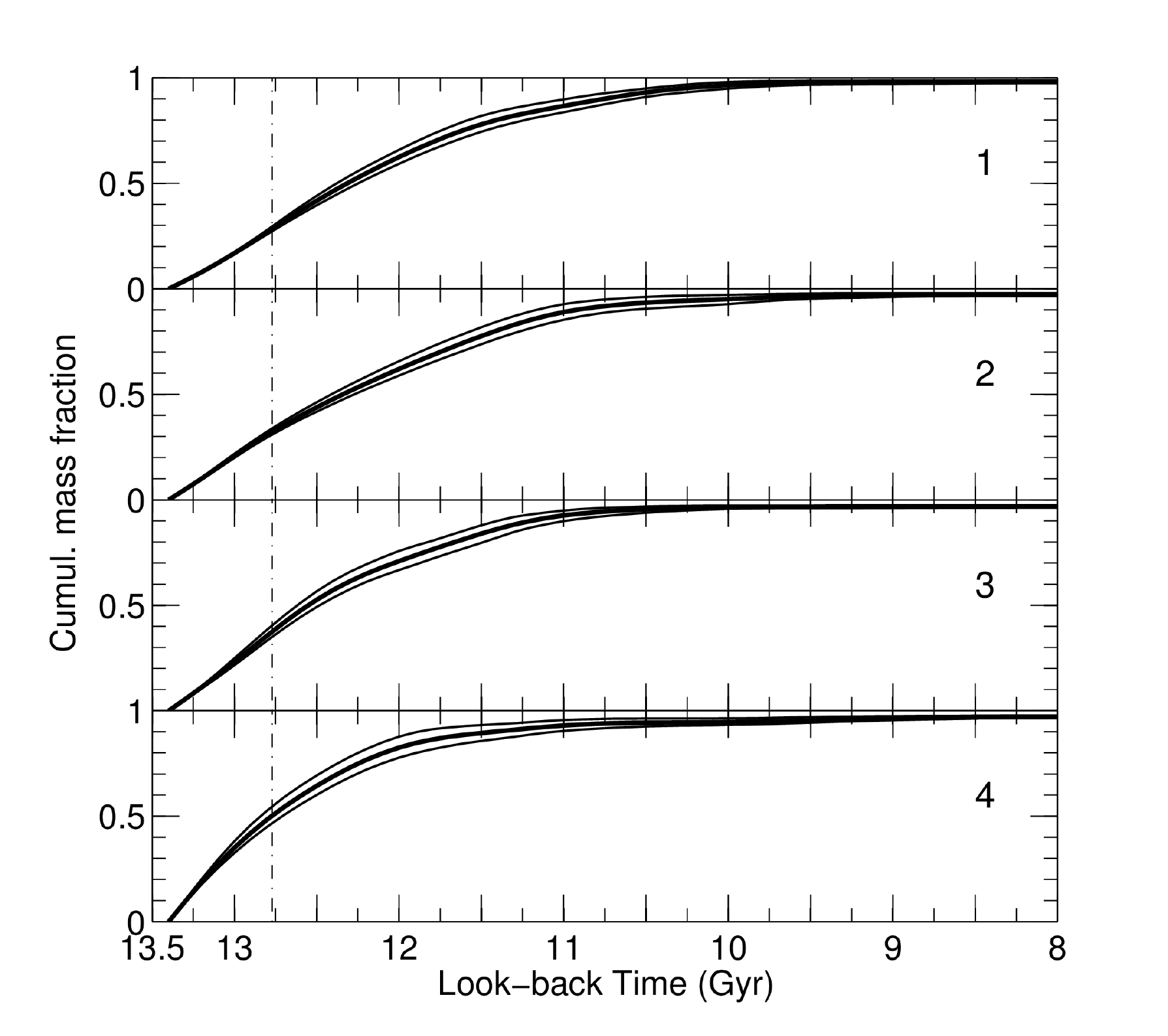}

   \caption[Resulting radial SFH, metallicity as a function of the look-back time and cumulative mass fraction of Sculptor]{Results of the radial SFH (upper panels), metallicity as a function of look-back time (middle panels) and cumulative mass fraction (lower panels) for the four regions defined as in Figure \ref{fig:circless}. Thin lines represent the uncertainties. In the panels each region has been indicated with the corresponding number from $1$ to $4$.}
\label{fig:sfhtot}
\end{figure}

\begin{figure*}
\includegraphics[width=\columnwidth]{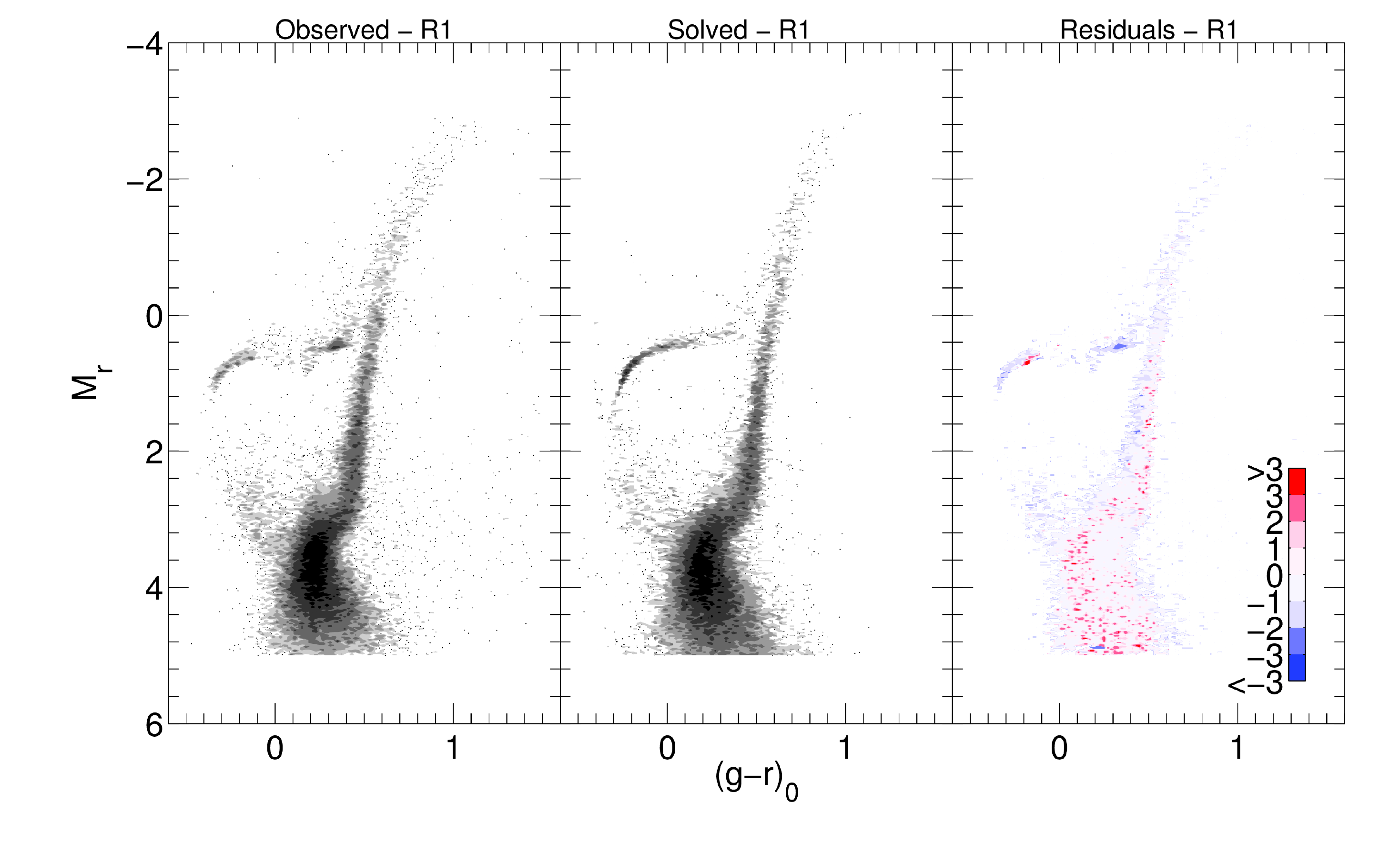}
\includegraphics[width=\columnwidth]{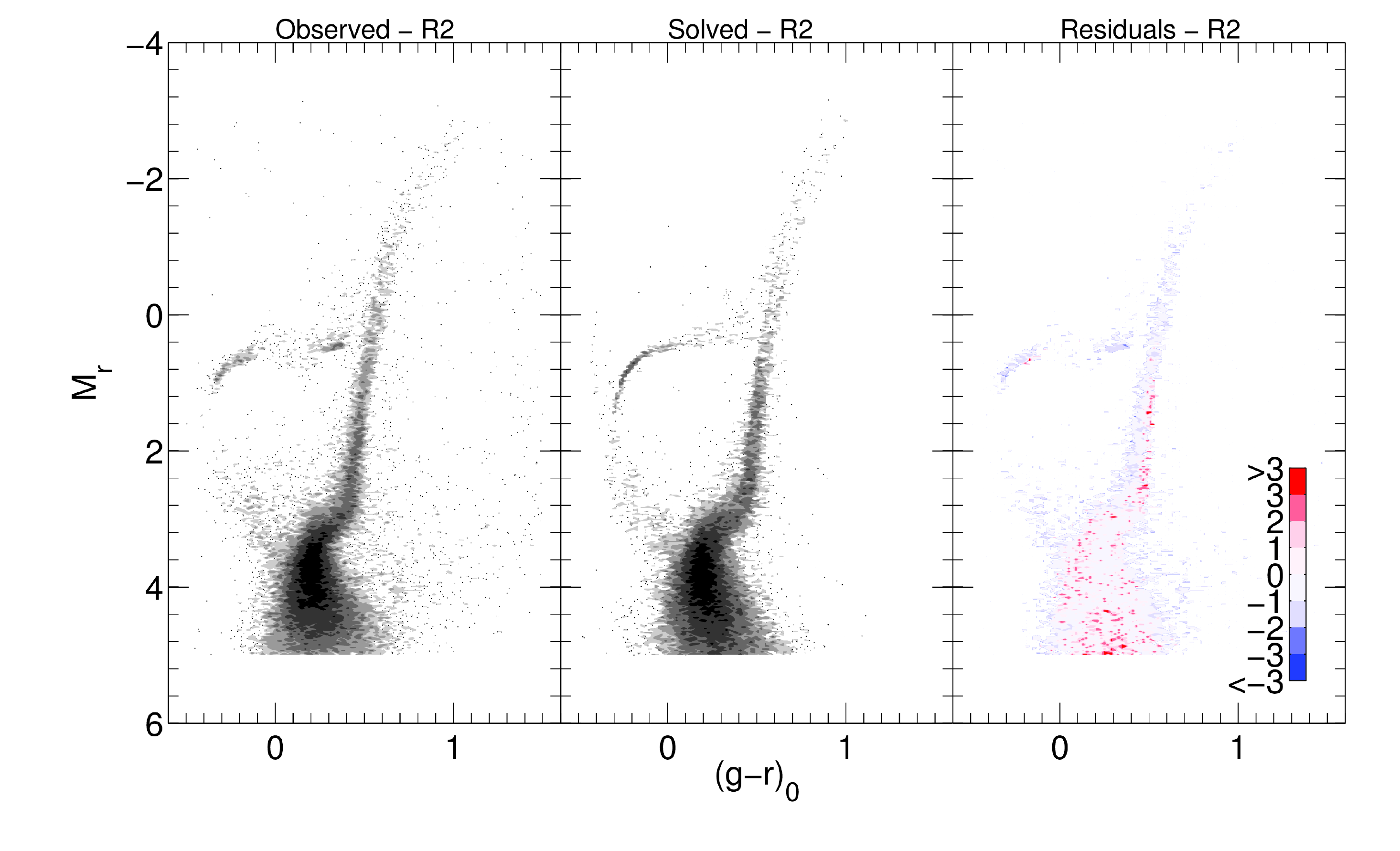}\\
\includegraphics[width=\columnwidth]{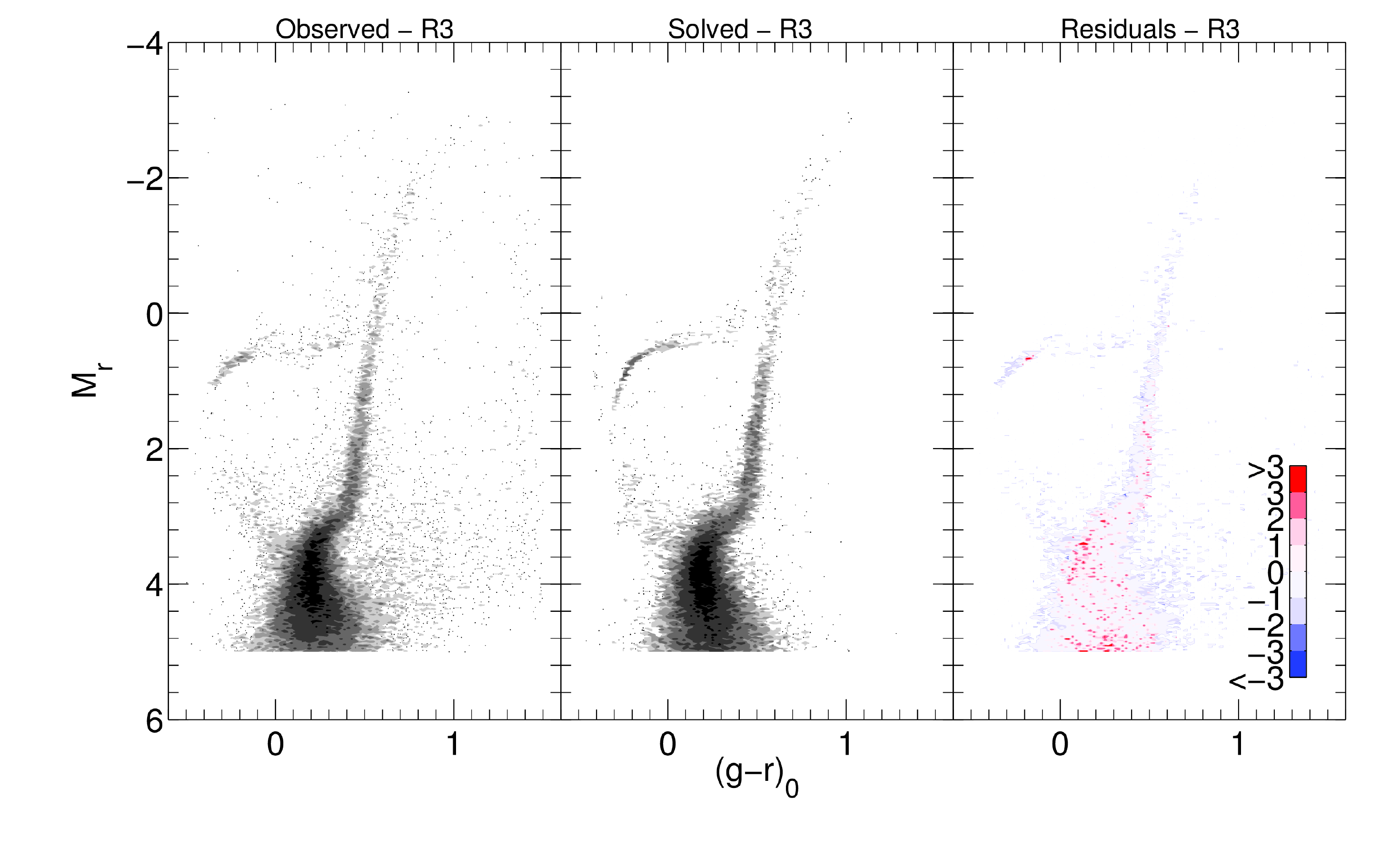}
\includegraphics[width=\columnwidth]{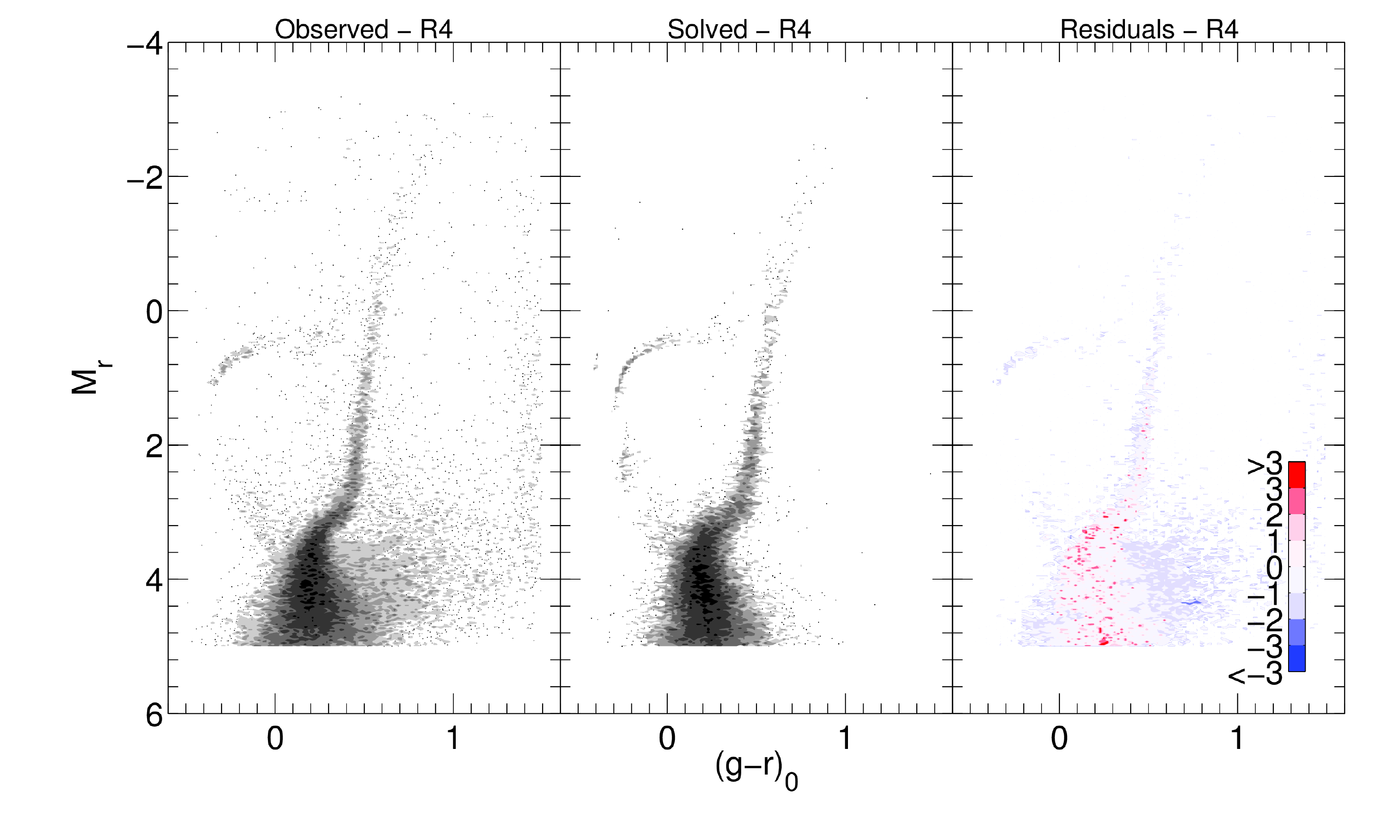}
   \caption{Upper left panel: as Figure \ref{fig:residuototale} but for stars within  $a\leq8.59$ arcmin (green points in Figure \ref{fig:circless}). Upper right panel: same as upper left panel but for stars within $8.59 <a\leq 13.41$ (magenta points in Figure \ref{fig:circless}). Bottom left panel: same as upper left panel but for stars within $13.41 <a\leq 19.35$ (blue points in Figure \ref{fig:circless}). Bottom right panel: same as upper left panel but for stars within $a\geq 19.35$ (red points in Figure \ref{fig:circless}).}
\label{fig:r1c}
\end{figure*}

In Fig.~\ref{fig:sfhtot} we show the resulting SFHs (upper panels), the metallicy as a function of time (middle panels) and the cumulative mass fraction (lower panels) for all the four regions and in Figure \ref{fig:r1c} the corresponding Hess diagrams of the oCMD (left panel), the best solution CMD (middle panel) and the residuals (right panel) are plotted. 
In each region the star formation is consistent, within our age resolution, with a single burst of star formation. From Fig.~\ref{fig:sfhtot} it is appreciable that the duration of star formation is shorter going outwards and we will constrain it for each case subsequently.
Within errors we have not detected any metallicity gradient. Also the mean ages, within $1 \sigma$ are consistent for the inner $3$ regions except region $4$ since the measured values are: $12.05\pm0.13$, $12.07\pm0.13$, $12.23\pm0.13$ and $12.45\pm0.14$ Gyr. By inspecting Fig.~\ref{fig:sfhtot} it can be noted that there is a trend of the SFH with distance that involves also the 3 innermost regions.
This suggests that a population gradient is present in Sculptor.

\subsection{Constraining the duration of the main SFH burst}
It is known that observational effects tend to broaden the main features of the SFH, mostly those reflecting the oldest episodes of star formation \citep{2016ApJ...823....9A}. To investigate this effect in the radial SFH, we have followed the procedure outlined in \citet{2011ApJ...730...14H}, to see if hints of different time scales of star formation are detectable.
For this test we derived the SFHs for a number of mock stellar populations in the same way as for observational data adopting for each radial region the corresponding model which accounts for the local photometric errors of the region under consideration.
The mock stellar populations are characterized by an increasing age width all starting $13.5$ Gyr ago. Even though the peak is situated at $\sim 12.8$ Gyr in three of the four regions, we use the oldest age so that the duration of the SFH obtained is a limit since it is expected a lower difference between the size of the injected and recovered SFH at younger ages.
The duration of each burst, in terms of Full Width at Half Maximum ($FWHM_{in}$), has been chosen to be: $0.25$, $0.5$, $0.75$, $1$, $1.5$, $2$ Gyr.
Metallicity has been fixed to $[Fe/H]=-1.86$, which is the mean metallicity recovered for Sculptor.

In Table~\ref{tab:numbers} we summarize for all the regions the mean recovered $FWHM_{rec}$ of each mock burst.

\begin{table*}
\centering
\footnotesize
\caption{List of the recovered SFH FWHM for the mock bursts in each elliptical region.}
\label{tab:numbers}
\begin{tabular}{ccccc} 
\hline
$FWHM_{in}$ & $FWHM_{rec-1}$ & $FWHM_{rec-2}$ & $FWHM_{rec-3}$ & $FWHM_{rec-4}$\\
\hline
$0.25$ & $2.98$ & $2.88$ &  $2.57$ & $2.28$\\
$0.5$ & $3.02$ & $2.92$ & $2.69$ & $2.35$\\
$0.75$ & $3.13$ & $2.98$ &  $2.87$ & $2.50$\\
$1$ & $3.37$ & $3.02$ &  $2.92$ & $2.55$\\
$1.5$ & $3.63$ & $3.09$ &  $3.28$ & $2.83$\\
$2$ & $4.00$ & $3.52$ &  $3.35$ & $2.96$\\
\hline
\end{tabular}
\end{table*}

\begin{figure*}

\includegraphics[width=\columnwidth]{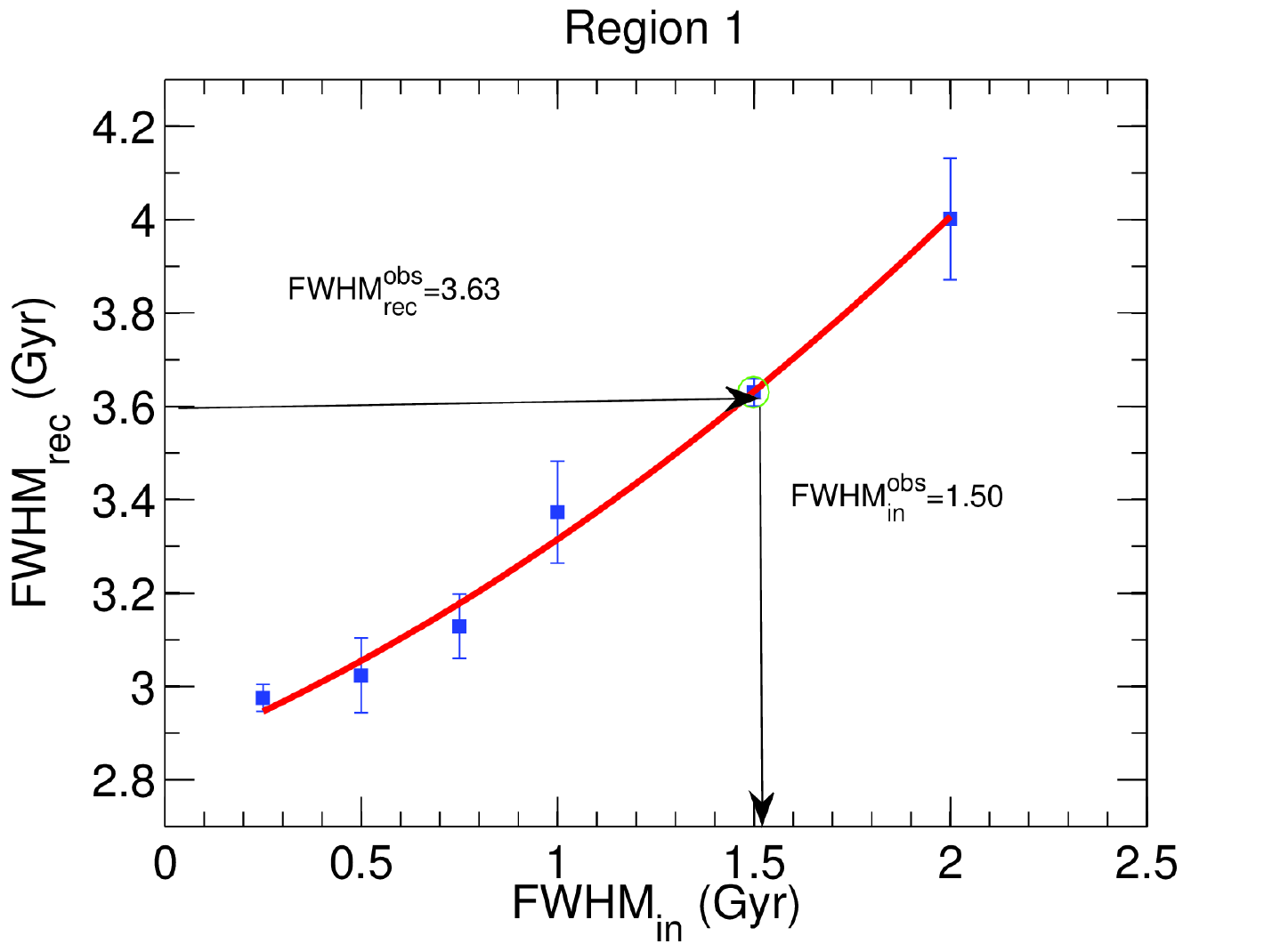}
\includegraphics[width=\columnwidth]{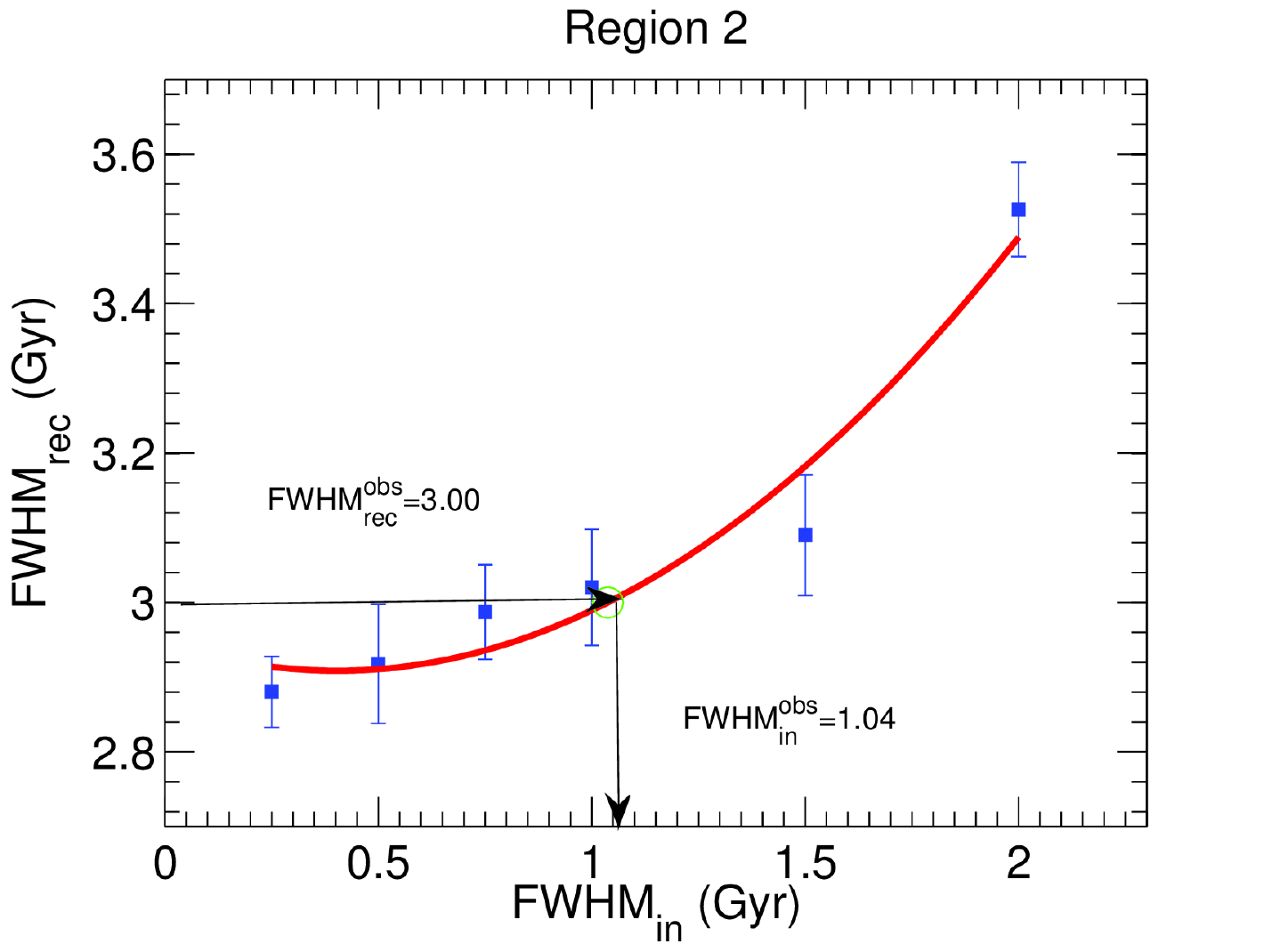}\\
\includegraphics[width=\columnwidth]{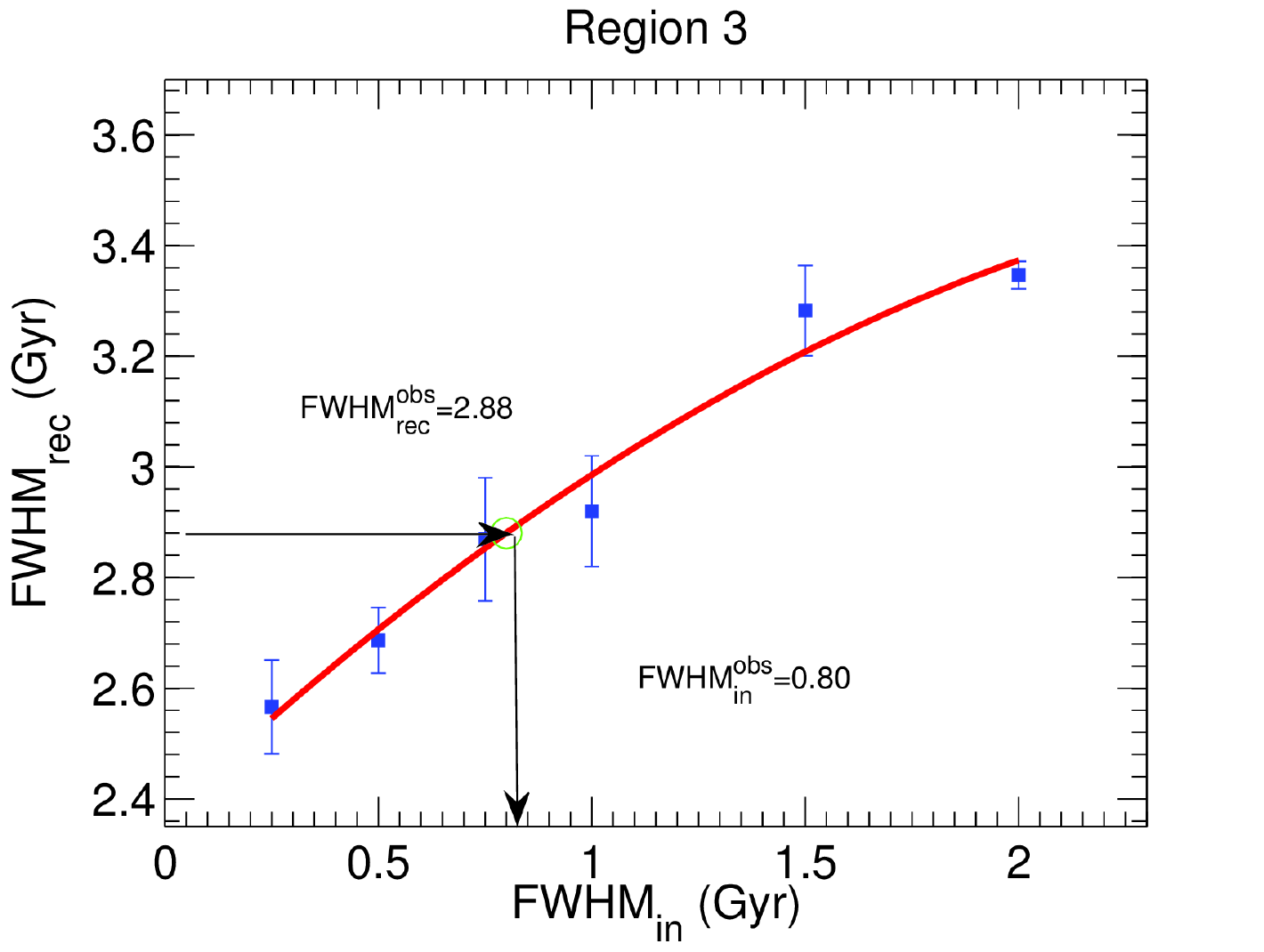}
\includegraphics[width=\columnwidth]{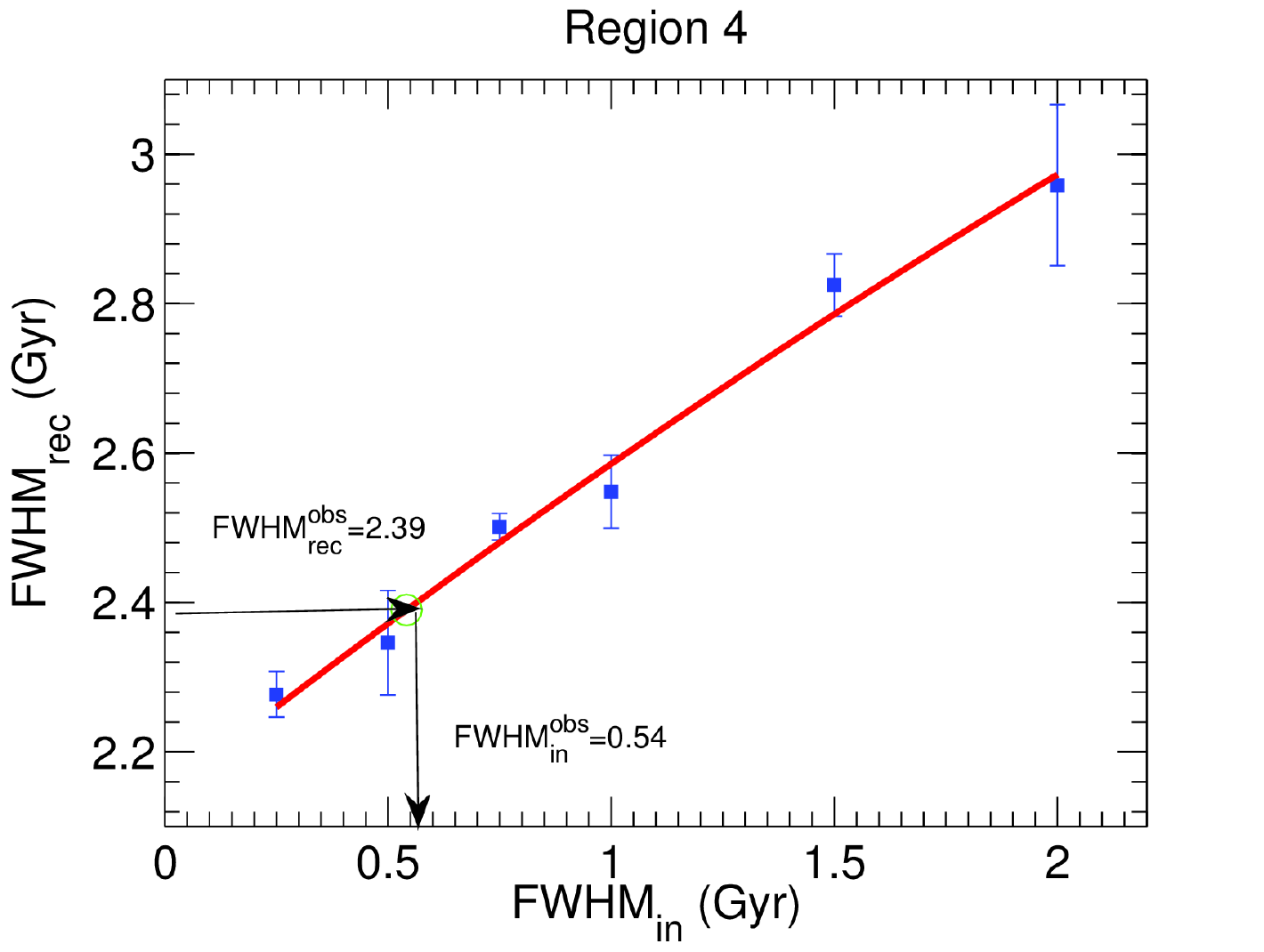}

   \caption{Burst duration confinement. Input FWHM of the mock bursts, $FWHM_{in}$, versus the recovered SFH FWHM, $FWHM_{rec}$ (blue dots). Points have been fitted with a quadratic polynomial (red line). Confinement of the star formation burst duration has been obtained by intercepting the fitting the red line (green circle) at the $FWHM_{in}$ value corresponding to the best solution SFH.
   Upper left panel: confinement relative to region $1$, the star formation burst duration results $FWHM_{in}^{obs}\sim1.5$ Gyr. Upper right panel: same as upper left panel but for region $2$; we confine the star formation event to a value of $FWHM_{in}^{obs}\sim1$ Gyr. Bottom left panel: same as upper left panel but for region $3$; we confine the star formation event to a value of $FWHM_{in}^{obs}\sim0.8$ Gyr. Bottom right panel: same as upper left panel but for region $4$; we confine the star formation event to a value of $FWHM_{in}^{obs}\sim0.54$ Gyr.}
\label{fig:limits1}
\end{figure*}

\begin{figure*}

\includegraphics[width=15 cm]{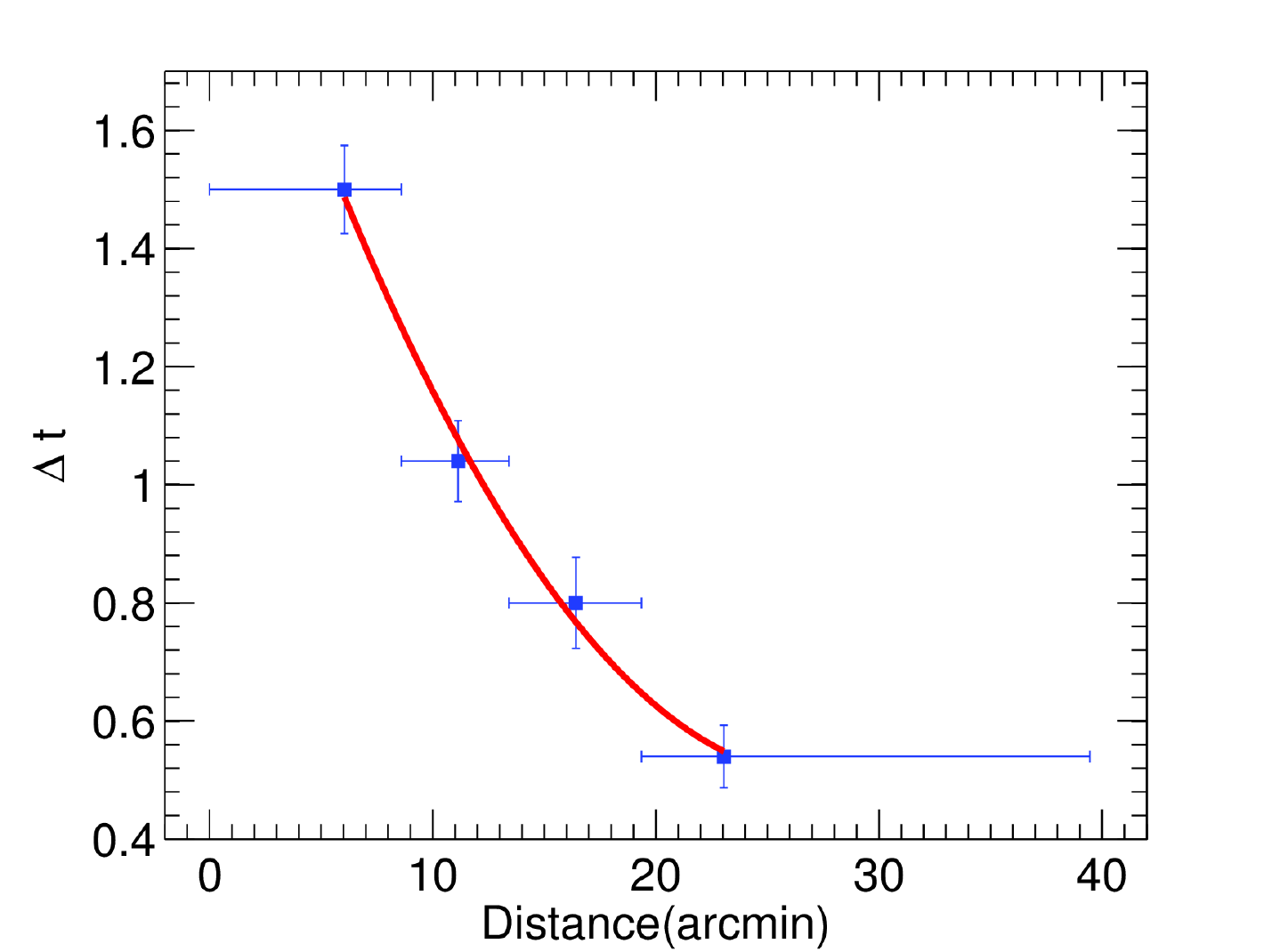}
   \caption{This figure summarizes the results achieved. The points represents the duration of the star formation in each region and a function of the distance from the center of Scultor. The horizontal error bars refer to the radial coverage of each individual region.}
\label{fig:riassumendo}
\end{figure*}

Fitting a single Gaussian profile to each SFH of Sculptor in the four regions we estimate the following typical width for the observed $\psi(t)$ in the age range $10-13.5$ Gyr: $\sigma_{1-4}=1.54, 1.28, 1.23, 1.02$ Gyr, which  corresponds to a $FWHM_{obs}=3.63, 3.01, 2.88, 2.39$ Gyr. 
In Figure~\ref{fig:limits1}  we have plotted, for each region, the $FWHM_{in}$ Gyr of the mock bursts and their associated mean recovered $FWHM_{rec}$. Fitting the resulting data with a quadratic polynomial we are able to constrain the period of star formation for each region. Table~\ref{tab:numbers1} summarizes the results, while Figure~\ref{fig:riassumendo} shows the constrained duration of the bulk of the star formation as a function of radius.
\begin{table}
\centering
\footnotesize
\caption{Constrainment of the period of star formation for each elliptical region. $FWHM_{rec}$ refers to the real measured $FWHM$ for each SFH, while $FWHM_{in}$ is the associated constrained burst found by means of the quadratic fit. We also provide the errors associated to $FWHM_{in}$ in the last column.}
\label{tab:numbers1}
\begin{tabular}{cccc} 
\hline
Region & $FWHM_{rec}$ & $FWHM_{in}$ & $\sigma FWHM_{in}$\\
\hline
$1$ & $3.63$ & $1.50$ & $0.07$\\
$2$ & $3.01$ & $1.04$ & $0.07$\\
$3$ & $2.88$ & $0.80$ & $0.07$\\
$4$ & $2.39$ & $0.54$ & $0.05$\\

\hline
\end{tabular}
\end{table}

The main star formation episode in each region decreases from the center outwards.
The star formation in region $1$ lasted $\sim 1.5\,\pm\,0.1$ Gyr, with its peak at $\sim 12.8$ Gyr ago. Its completion is well after the end of the epoch of reionization, which is fixed at $\sim 12.77$ Gyr \citep{2001AJ....122.2850B}. 
The star formation episode in region $4$ has been confined to  $\sim 0.5$ Gyr. Since the peak of star formation is at $\sim 13.5$ Gyr ago, in this region star formation terminated before the end of the epoch of reionization.
This suggest that Sculptor, unlike Sextans, has continued forming stars in its inner regions, because enriched gas probably concentrated there and permitted a prolonged period of star formation. 

However, the peculiarity of Sculptor is the fact that it is embedded in two symmetrical HI clouds. It is thus interesting to investigate if this neutral gas could have been ejected by SN winds \citep{1998AJ....116.1690C}. 
To this aim, the mechanical luminosity of the SNe released during the main star formation episode has been calculated as outlined in \citet{2011ApJ...730...14H}. We then compared this quantity, along with the value of the mass of gas of the galaxy, with the results presented in \citet{1999ApJ...513..142M}. This in order to infer if Sculptor had lost mass in the past and in the case the modality according to which it happened.

From the obtained SFH results that a total of $2.24\times10^{6}\,{\rm M}_{\odot}$ of gas was converted into stars. This value has been obtained scaling the resulting SFH to the whole galaxy using the King profile of \citet{2006AJ....131..375W}.
Assuming a minimum progenitor mass for core collapse SNe of $6.5\,{\rm M}_{\odot}$ \citep{2005essp.book.....S}, a total of $3.79\times10^{4}$ SNe have been obtained.
In the case of a progenitor minimum mass of $10\,{\rm M}_{\odot}$, it results instead $2.12\times10^{4}$ SNe.
For the two mass values for type II SNe listed above, we calculated a total mechanical luminosity released during the star formation episode of $1.5$ Gyr of $L_{w}=8.01\times10^{38}$ erg/s and $L_{w}=4.48\times10^{38}$ erg/s respectively. This was calculated assuming an energy release of $10^{51}$erg per SN \citep{1999ApJS..123....3L}. 

These values would place Sculptor in the region of blow-out/mass loss regime, from the comparison with the model results of \citet{1999ApJ...513..142M} shown in their Figure 1. 
In a "blowout" the central SN explosions blow a hole through the intergalactic medium, accelerating some fraction of gas and releasing the energy of subsequent explosions without major effects on the remaining gas. 

In other words, the star formation began with an initial strong burst slowed by gas loss due to supernovae explosions. Later the remaining gas concentrated again in the innermost region permitting a second or more episodes of star formation. 
The confinement obtained for the star formation episode of region $4$ is likely the record of the primordial large scale star formation of Sculptor, which corresponds to the BHB stars that are present in each partial CMD in Figure \ref{fig:hk}.

Knowing the radial SFHs and following the prescription in \citet{2013ApJ...778..103H} we are able to provide some predictions about the  structural properties of this dwarf, such as the core radius and the total stellar mass. We fit our results with an exponential profile, $\psi(r)=M_{0}e^{-r/{\alpha_{\psi}}}$, where $M_{0}$ is the central mass density and $\alpha_{\psi}$ is the scale length\footnote{radius at which mass drops by $e^{-1}$}. For Sculptor we derived $\alpha_{\psi}=198 \pm 6$ pc and a core radius $CR_{\psi}=320 \pm 6$ pc. Under the assumptions that the observed field provides a good representation of the whole galaxy and the radial profile of the stellar mass density follows the same $\psi(r)$ beyond the observed radius, we integrated the exponential profile, deriving a total stellar mass of $M_{\*}=(6.3 \pm 0.4) \times 10^{6} M_{\odot}$. These values are very similar to the ones obtained for Cetus in \citet{2013ApJ...778..103H}. This fact can indicate that the two galaxies have followed similar evolutionary paths.


\section{Discussion}\label{sec:discsculptor}
It is interesting to compare the results presented above with earlier work, such as the star formation computed by \citet{2012A&A...539A.103D}, who have found an extended, continuous star formation of $6-7$ Gyr. Our results show that Sculptor stopped forming stars $\sim 11.3$ Gyr ago, thus indicating a duration of $\sim 2.2$ Gyr. The difference could be explained if we consider that our photometry is deeper that the one presented in \citet{2011A&A...528A.119D} and allows us to better sample the oldest MS TO.
From their radial study, \citet{2012A&A...539A.103D} found that the innermost annulus (whose extension is quite similar to the one adopted in the present analysis) can not be modeled with a single narrow burst of star formation, while, for the outer annulus of their catalogue, the recovered SFH can be well approximated to a single short burst of star formation.
As \citet{2012A&A...539A.103D} discussed, it is particularly challenging to properly reproduce the innermost region of Sculptor, in part for crowding effects but mostly for the complex star formation.
This fact can be intuited also inspecting Fig.~\ref{fig:hk}: all the features of the CMD appear more narrow going outwards. The broadening can be attributed to metallicity/age spread, thus suggesting a longer star formation in the
central regions.

From our analys we derived a mean metallicity for Sculptor of $[Fe/H]=-1.86$ . This value is in accordance with the mean value measured by means of spectroscopy of single stars by \citet{2006ApJ...651L.121H} along a radius of $40$ arcmin from the center of Sculptor. \citet{2004ApJ...617L.119T} identified two distinct stellar components in Sculptor, one metal-rich $-0.9>[Fe/H]>-1,7$, one metal-poor $-1.7>[Fe/H]>-2.8$. The peak of the global MDF is centered at $[Fe/H]=-1.8$.
Other spectroscopic measurements have been perfomed by \citet{2009ApJ...705..328K} who measured a mean metallicity of $[Fe/H]=-1.58$ and by \citet{2008MNRAS.383..183B} that found a mean of $[Fe/H]=-1.56$, both sampled stars out to about $11$ arcmin.

From the constraining of the SFHs radially we derived a period of $\sim 0.5$ Gyr for the outermost region which is in striking agreement with the single early episode of $0.5$ Gyr computed by \citet{2014ApJ...782L..39A} during which SNe II explosions would be sufficient to build a substantial dark matter core.

The radial gradient found in Sculptor is in agreement with outside-in scenarios of dwarf galaxy evolution, which has been found in other dwarf galaxies of the Local Group \citep{2013ApJ...778..103H}, when these type of galaxies run out of gas on the outskirts but are able to keep forming stars in the center for a longer period.

Finally, we would like an answer to the question if Sculptor is a fossil of the pre-reionization era, as introduced in \citet{2005ApJ...629..259R}: a dwarf that has experienced more than the $70\%$ of its  star formation before the end of the reionization and that has a luminosity $L_V<10^6 \,L_\odot$.
Sculptor does not satisfies the second condition, since it has a luminosity $L_V=(2.03\pm0.79)\times 10^6 \,L_\odot$ \citep{2009MNRAS.394L.102L}.   
According to our results region $1$ is characterized by a duration of the episode of star formation of $\sim 1.5$ Gyr which indicates a star formation extended also after the end of the epoch of reionization. For the outermost region $4$, on the contrary, its star formation has been confined to $\sim 0.5$ Gyr. There the star formation ended before or contemporaneous with the end of the epoch of reionization.
From the considerations listed above, Sculptor can not be classified strictly as a true fossil of the pre-reionization era as opposed to what we found for Sextans \citep{2018MNRAS.476...71B}. 
Our results suggest that Sculptor could have suffered in the past a blowout as described in \citet{1999ApJ...513..142M}. It is thus tempting to link this to the possible formation of the HI distribution embedding Sculptor \citep{1998AJ....116.1690C}. 

\section{Summary and Conclusions}\label{sec:concsculptor}
We have derived the global and radial SFH for the Sculptor dSph based on deep $g$,$r$ photometry taken with DECam at the Blanco telescope.

The age resolution of our derived SFH indicates that Sculptor has experienced a single event of star formation limited to the first $\sim 2$ Gyrs after Big Bang, producing $\sim 70 \%$ of its mass about $12$ Gyr ago.
The mean metallicity retrieved is $[Fe/H]\sim \,-1.8$.
The retrieved metallicities are consistent with the spectroscopic measurements by  \citet{2006ApJ...651L.121H} and \citet{2004ApJ...617L.119T}. 
We have investigated how the SFH of Sculptor changes radially, subdividing in four regions the sampled area. In each region the star formation is consistent, within our age resolution, with a single burst of star formation of different duration. We find that the duration of the episodes of star formation increases towards the centre and we provide the intrinsic duration of these bursts. The innermost region presents the longer period of star formation of $\sim 1.5$ Gyr, in agreement with the estimate by \citet{2009ApJ...705..328K} via chemo-dynamical models, going outwards it decreases to $\sim 0.54$ Gyrs. 
These results suggest that Sculptor continued forming stars after the reionization epoch in its central part, while in the peripheral region the majority of stars were formed before or coincident with the end of the reionization epoch.
Our results are compatible with an outside-in scenario of dwarf galaxy formation.
Finally, from the calculation of the mechanical luminosity produced by SNe we can advance the hypothesis that Sculptor has suffered a 'blow out' in its early epochs that does not completely inhibit star formation. This result, together with the constraints by \citet{2005ApJ...629..259R}, indicates that Sculptor can not be strictly considered a fossil of the pre-reionization era.

\section*{Acknowledgements}
We thank the anonymous referee for the pertinent comments and suggestions that have helped us to improve this paper.
MB, SLH, SC, AA and GP  acknowledge support from the Spanish Ministry of Economy and Competitiveness (MINECO) under grant AYA2013-42781.

This research uses services or data provided by the Science Data Archive at NOAO. NOAO is operated by the Association of Universities for Research in Astronomy (AURA), Inc. under a cooperative agreement with the National Science Foundation.
This project used data obtained with the Dark Energy Camera (DECam), which was constructed by the Dark Energy Survey (DES) collaboration. Funding for the DES Projects has been provided by the U.S. Department of Energy, the U.S. National Science Foundation, the Ministry of Science and Education of Spain, the Science and Technology Facilities Council of the United Kingdom, the Higher Education Funding Council for England, the National Center for Supercomputing Applications at the University of Illinois at Urbana-Champaign, the Kavli Institute of Cosmological Physics at the University of Chicago, the Center for Cosmology and Astro-Particle Physics at the Ohio State University, the Mitchell Institute for Fundamental Physics and Astronomy at Texas A\&M University, Financiadora de Estudos e Projetos, Funda{\c c}{\~a}o Carlos Chagas Filho de Amparo {\`a} Pesquisa do Estado do Rio de Janeiro, Conselho Nacional de Desenvolvimento Cient{\'i}fico e Tecnol{\'o}gico and the Minist{\'e}rio da Ci{\^e}ncia, Tecnologia e Inovac{\~a}o, the Deutsche Forschungsgemeinschaft, and the Collaborating Institutions in the Dark Energy Survey. 
The Collaborating Institutions are Argonne National Laboratory, the University of California at Santa Cruz, the University of Cambridge, Centro de Investigaciones En{\'e}rgeticas, Medioambientales y Tecnol{\'o}gicas-Madrid, the University of Chicago, University College London, the DES-Brazil Consortium, the University of Edinburgh, the Eidgen{\"o}ssische Technische Hoch\-schule (ETH) Z{\"u}rich, Fermi National Accelerator Laboratory, the University of Illinois at Urbana-Champaign, the Institut de Ci{\`e}ncies de l'Espai (IEEC/CSIC), the Institut de F{\'i}sica d'Altes Energies, Lawrence Berkeley National Laboratory, the Ludwig-Maximilians Universit{\"a}t M{\"u}nchen and the associated Excellence Cluster Universe, the University of Michigan, {the} National Optical Astronomy Observatory, the University of Nottingham, the Ohio State University, the OzDES Membership Consortium, the University of Pennsylvania, the University of Portsmouth, SLAC National Accelerator Laboratory, Stanford University, the University of Sussex, and Texas A\&M University.

Based on observations at Cerro Tololo Inter-American Observatory, National Optical Astronomy Observatory, which is operated by the Association of Universities for Research in Astronomy (AURA) under a cooperative agreement with the National Science Foundation.

This research used the facilities of the Canadian Astronomy Data Centre operated by the National Research Council of Canada with the support of the Canadian Space Agency. 
This research has made use of the NASA/IPAC Extragalactic Database (NED) which is operated by the Jet Propulsion Laboratory, California Institute of Technology, under contract with the National Aeronautics and Space Administration. 












\bsp	
\label{lastpage}
\end{document}